\documentclass[useAMS,usenatbib]{mn2e}

\newcommand{\teff}{\mbox{$T_{\rm eff}$}}

\newcommand{\msun}{M\ensuremath{_\odot}}

\newcommand{\degree}{\mbox{\ensuremath{^\circ}}}   

\usepackage{color}
\usepackage{epsfig}
\usepackage{psfig}
\usepackage{lscape}
\usepackage{supertabular}
\usepackage{graphicx}
\usepackage{chngpage}

\title[Lucky Imaging of transiting planet hosts]{Lucky Imaging of transiting planet host stars with LuckyCam}

\author[F. Faedi et al.]
{F.~Faedi$^{1}$\thanks{E-mail: f.faedi@warwick.ac.uk -- Part of this work was carried out while at Queens University Belfast ${^7}$},
 T.~Staley$^2$, Y.~G\'omez~Maqueo~Chew$^{3,1}$, D.~Pollacco$^{1}$, 
 \newauthor S.~Dhital$^{3}$, S.~C.~C.~Barros$^{5}$, I.~Skillen$^6$, L.~Hebb$^3$, C.~Mackay$^2$, and C.\ A.~Watson$^7$\\
$^1$Department of Physics, University of Warwick, Coventry CV4 7AL, UK\\
$^2$Institute of Astronomy, Cambridge University, Madingley Road, Cambridge CB3 0HA, UK\\
$^3$Department of Physics and Astronomy, Vanderbilt University, Nashville, TN 37235, USA\\
$^4$Department of Astronomy, Boston University, 725 Commonwealth Avenue, Boston, MA 02215, USA \\
$^5$Aix Marseille Universit\'e, CNRS, LAM (Laboratoire d'Astrophysique de Marseille) UMR 7326, 13388, Marseille, France\\
$^6$Isaac Newton Group of Telescope, Apartado de Correos 321, Santa Cruz de la Palma, 38700, Spain \\
$^7$Astrophysics Research Centre, Queen's University Belfast, University Road, Belfast BT7 1NN, UK }

\begin{document}

\date{}


\maketitle

\label{firstpage}

\begin{abstract}
  We obtained high-resolution, high-contrast optical imaging in the
  SDSS~$i'$ band with the LuckyCam camera mounted on the 2.56m Nordic
  Optical Telescope, to search for faint stellar companions to 16
  stars harbouring transiting exoplanets. The Lucky Imaging technique
  uses very short exposures to obtain near diffraction-limited images
  yielding sub-arcsecond sensitivity, allowing us to search for faint
  stellar companions within the seeing disc of the primary planet
  host. Here we report the detection of two candidate stellar
  companions to the planet host TrES-1 at separations $<6.5\arcsec$
  and we confirm stellar companions to CoRoT-2, CoRoT-3, TrES-2,
  TrES-4, and HAT-P-7 already known in the literature. We do not
  confirm the candidate companions to HAT-P-8 found via Lucky Imaging
  by \citet{Bergfors2013}, however, most probably because HAT-P-8 was
  observed in poor seeing conditions.  Our detection sensitivity
  limits allow us to place constraints on the spectral types and
  masses of the putative bound companions to the planet host stars in
  our sample.  If bound, the stellar companions identified in this
  work would provide stringent observational constraints to models of
  planet formation and evolution. In addition these companions could
  affect the derived physical properties of the exoplanets in these
  systems.
  
 \end{abstract}

\begin{keywords}
  instrumentation: high angular resolution -- methods: observational
  -- stars: binaries --stars: planetary systems
\end{keywords}

\section{Introduction}
More than 800 extrasolar planets have been discovered to date showing
a large variety of physical and dynamical properties that are
dramatically different from those observed in our Solar System. This
has revolutionised our understanding of planetary formation, structure
and evolution. One third of the known gas giant planets orbit their
host at separations smaller than a few tenths of an AU (with orbital
periods ${\rm P}<10$~d). Among these, transiting systems are specially
important as they allow accurate measurements of masses, radii, and
hence densities, to be derived.  These key parameters inform us of the
system's physical properties, and can constrain theoretical
evolutionary models (e.g.  \citealt{Guillot2005};
\citealt{Fortney2007}; \citealt{Burrows2007}; \citealt{Liu2008}). In
contrast to the planets in our Solar system, exoplanets show a large
variety of orbital properties, for example their orbital
eccentricities span a wide range $e$ = 0--0.97 (e.g. HD\,80606,
\citealt{Pont2009} and \citealt{Eggenberger2004}; HAT-P-13,
\citealt{Bakos2009}). The close-in ``hot Jupiters'' show a large
angular distribution of (mis)alignments with respect to their host
stars' rotation axis (\citealt{Winn2010b,Winn2011},
\citealt{Triaud2010}, \citealt{Morton2011}), and some exoplanets even
have retrograde orbits (e.g. WASP-17, \citealt{Anderson2010}).

To explain the observed exoplanet orbital configurations, different
scenarios have been proposed for migrating the planets inward
from beyond the snow line to their observed position. These migration
mechanisms make different predictions about the current orbital
configurations of the planetary systems. For example, planet-disc
interaction via angular momentum exchange (e.g. \citealt{Lin1996}, and
\citealt{Ida2004}) results in damping any initial inclination of the
planetary orbit with respect to the disc (see e.g.,
\citealt{Marzari2009,Watson2011}).  Alternatively, gravitational interaction
among multiple giant planets (planet-planet scattering;
e.g. \citealt{Wu2003}, \citealt{Nagasawa2008}), and perturbations
induced by a companion star or a more distant massive planet (Kozai
mechanisms, see \citealt{Fabrycky2007}) result in orbital
configurations with large spin-orbit misalignments and large eccentricities.
(\citealt{Rasio1996}; \citealt{Weidenschilling1996},
\citealt{Chatterjee2008}).
 
Observational evidence for planet-disc migration is found in
multi-planetary systems with mean-motion resonant orbits (e.g. GJ 876,
\citealt{Lee2002, Crida2008}).  On the other hand, measurements of the
Rossiter-Mclaughlin effect\footnote{See the Holt--Rossiter--McLaughlin
  Encyclopaedia;
  http://www.aip.de/People/rheller/content/main\_spinorbit.html}
(\citealt{Rossiter1924}; \citealt{McLaughlin1924}) suggest that
$\sim$40\% of transiting planets have highly tilted orbits providing
supporting evidence for planet-planet scattering and the Kozai
migration mechanism (\citealt{Winn2009b}; \citealt{Winn2010a}).
Examples of systems with large spin-orbit misalignments and/or high
eccentricities are respectively WASP-17b, \citep{Anderson2010},
HAT-P-7b \citep[e.g.,][]{Winn2009a}, and HD80806b
\citep[e.g.,][]{Pont2009,Eggenberger2004,Hebrard2010}.  More recently
\citet{Albrecht2012} suggested that the Kozai mechanism is responsible
for the migration of the majority, if not all, hot Jupiters, both
mis-aligned and aligned, and that star-planet tidal interaction plays
a central role in shaping exoplanets orbital configurations.

In this paper we present high-contrast, high-angular resolution
optical imaging for 16 stars harbouring transiting extrasolar planets
to search for faint stellar companions. Identifying binary companions
to known planet hosts can provide observational evidence to constrain
the different formation and evolution scenarios, as well as provide
crucial information for subsequent exoplanet characterisation
\citep[see also][]{Bergfors2013,Daemgen2009,Narita2012}. The presence
of a close-in stellar source to a transiting planet host star, as in
the case for WASP-12 \citep[via Lucky Imaging]{Bergfors2013}, could
affect the derived planetary parameters by diluting the transit signal
\citep[see also][]{Daemgen2009}.  For example, \citet{Crossfield2012}
find that WASP-12b is rather hotter and slightly larger (by 1--2\%)
than previously reported, highlighting the importance of
high-resolution imaging for the characterisation of known and newly
discovered transiting planetary systems. Additionally, the presence of
an M dwarf only 1\arcsec~from WASP-12 might have contaminated past
atmospheric measurements, possibly challenging the detection of a high
atmospheric C/O ratio for WASP-12b (\citealt{Madhusudhan2011}, and
\citealt{Crossfield2012}, for a recent re-analysis).

The paper is organised as follows: in \S2 we briefly describe our
Lucky Imaging technique; \S3 presents our LuckyCam observations; in
\S4 we explain the data reduction, image analysis and candidate
detection.  Our results are presented in \S5, including the
non-detections in our sample.  In \S6, we discuss the likelihood of
the detected companions being bound to the planet hosts. Finally, we
summarise our findings and conclusions in \S7.

\begin{figure*}
\centering
 \psfig{file=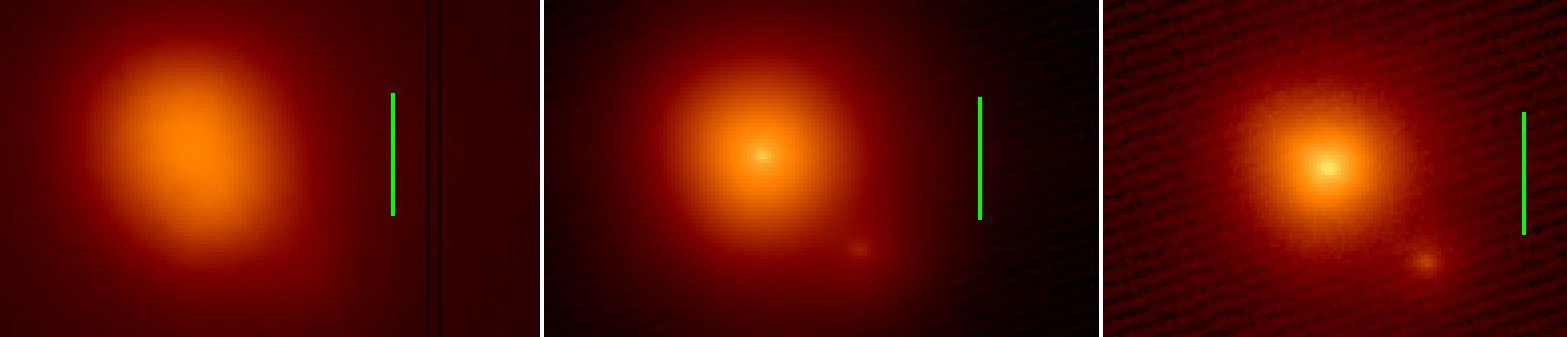,width=0.95\textwidth}
 \caption{Images of TrES-2 obtained with LuckyCam showing the image
   quality improvement due to the Lucky Imaging technique.  Left-hand
   panel is a simple average corresponding to a conventional long
   exposure. The middle and right-hand panels show images resulting
   from re-centering and drizzling the short exposures with 100 percent
   and 5 percent selection cutoffs, respectively.  All images have the
   same log scale. The green bar depicts 1\arcsec ~ line. Average
   seeing during these observations was 0.8\arcsec.  }
\label{fig1}
\end{figure*}

\section{Lucky Imaging Technique}
Lucky Imaging consists of the acquisition of short exposures, at a
rate of a few tens of frames per second, using a very low-noise
electron multiplying CCD camera (\citealt{Fried1978};
\citealt{Baldwin2001}; \citealt{Tubbs2002}; \citealt{Mackay2004};
\citealt{Law2006}). This allows the rapid image motion due to
atmospheric turbulence to be corrected. Because the perturbations
introduced by the atmosphere change on timescales of a few
milliseconds (known as the atmospheric coherence time), with fast
imaging each frame captures a different point spread function (PSF)
resulting from the atmospheric turbulence at that particular moment.
By monitoring the rapid PSF variations, we can select high quality
short exposures from moments of excellent seeing. During data
reduction the best frames are selected, aligned and co-added to
produce a final image with a bright diffraction limited core
surrounded by a fainter seeing halo. \citet{Law2006} give a detailed
explanation on the Lucky Imaging technique and the LuckyCam
specifications.

\section{Observations}

Observations were obtained between July 18 and July 22, 2009 at the
2.56m Nordic Optical Telescope (NOT) at the Roque de los Muchachos
Observatory, La Palma, with the Cambridge LuckyCam vistor
instrument. Seeing ranged from $\sim0.6\arcsec$ to $\sim1.65\arcsec$
as measured by the DIMM (at 500nm; \citealt{Tokovinin2002}). All
observations were made in the SDSS~$i'$ band, using a plate scale of
32.4~mas/pixel, providing good sampling of the PSF. The camera frame
rate was 20.75 frames per second using full chip readout (1024 pixels
squared). Table~\ref{tab:obs} presents a summary of our observations.
Targets were often observed slightly off-centre on the CCD detector to
achieve better positioning of the mosaic field of view for astrometric
calibration. The target observed closest to the CCD edge has an
unbroken observation area of radius 6.5\arcsec. Therefore, to give a
uniform dataset we only list detections within 6.5\arcsec.  However,
we note that the planet host HAT-P-1 \citep{Bakos2007} is known to be
part of a binary system with a companion at $\sim$11\arcsec, that was
clearly detected in our images at a separation $r =
11.26\pm0.03\arcsec$, although this target is not discussed further in
this paper.

We selected our sample to optimise the number of planet host stars
observable as by July 2009, in order to cover a large parameter space
of different stellar and planetary properties. Detailed information on
individual objects is available from the Exoplanet Encyclopaedia
\footnote{http://exoplanet.eu/}.  We present our sample in
Table~\ref{tab:obs}, separating the planet host stars with candidate
companions detected in this work from those without detections.

\begin{table*} 
  \caption{The sample of 16 stars harbouring transiting
    extrasolar planets studied in this paper. We list the spectral
    type and the orbital eccentricity ({\em e}) $^{2}$,  $\lambda$ the measured spin-orbit
    (mis)alignment angle (data and references taken from the Rossiter-McLaughlin encyclopaedia$^{1}$), 
    the number of observed frames and the total exposure times for the 100\% selection images of LuckyCam, and the average seeing at 500nm. }
\label{tab:obs}
\begin{tabular}{llccrlc}
\hline \hline 
Target 		&SpT &{\em e}&$\lambda$& N$_{\rm frames}$&T$_{\rm exp}$ &seeing    \\		
&&&(deg)&&(sec)&(\arcsec)	  \\
\hline
HAT-P-1		&G0V&0.067&$3.7\pm2.1$& 5400	&260	&0.65\\
HAT-P-2		&F8V&0.52& $1.2\pm13.4/ 0.2^{+12.2}_{-12.5} / 9\pm10$& 8000	&384	&0.99\\	
HAT-P-5		&G1V&0& $-$& 9000	&432	&1.02\\
HAT-P-6		&F8V&0&$166\pm10/165\pm6$& 5000	&240	&0.66 \smallskip \\
HAT-P-8$^a$	&F&0& $-9.7^{+9.0}_{-7.7}/-17^{+9.2}_{-11.5}$& 9100	&437	&1.51 \smallskip \\
HAT-P-11		&K4V&0.19&$103^{+22}_{-18} /103^{+26}_{-10} /106^{+15}_{-11} / 97_{-4}^{+8}$& 6000	&288	&0.64 \smallskip \\		
HD\,209458	&G0V&0&$3.9^{+18}_{-21}/ -4.4\pm1.4/ -5\pm7$& 10000	&480	&0.96 \smallskip \\
WASP-3		&F7V&0&$13^{+9}_{-7} / 3.3^{+2.5}_{-4.4} /5^{+6}_{-5} $& 10000	&480	&1.11\\
WASP-3		& ''&''& ''& 5000	&240	&0.65\\
WASP-10		&K5V&0.05& $-$& 5000	&240	&0.74\\
XO-1	        		&G1V&0& $-$& 10000	&480	&0.79\\
\hline
\multicolumn{6}{l}{Targets with candidate companions from this work} \\
\hline
CoRoT-2$^b$	&G7V&0&$7.2\pm4.5 / -1^{+6}_{-7.7} / 4.7\pm12.3 $& 6000	&288	&1.37 \smallskip \\
CoRoT-3$^c$	&F3V&0&$-37.6_{-10}^{+22.3}$& 5174	&248	&1.44 \smallskip \\
HAT-P-7$^d$	&F6V&0&$182.5\pm9.4/-132.6^{+10.5}_{-16.3} /155\pm37$& 5175	&248	&1.10\\
TrES-1		&K0V&0&$30\pm21$& 7700	&370	&0.88\\
TrES-2$^e$	&G0V&0&$-9\pm12$& 7000	&336	&0.83\\
TrES-4$^e$	&F8V&0&$6.3\pm 4.7$& 10000	&480	&1.12\\
\hline
\multicolumn{6}{l}{$^{a}$ Candidate companion identified by \citet{Bergfors2013}} \\
\multicolumn{6}{l}{$^{b}$ Candidate companion identified by \citet{Alonso2008}} \\
\multicolumn{6}{l}{$^{c}$ Candidate companion identified by \citet{Deleuil2008}} \\
\multicolumn{6}{l}{$^{d}$ Candidate companion identified by \citet{Narita2010}} \\
\multicolumn{6}{l}{$^{e}$ Candidate companion identified by \citet{Daemgen2009}} \\
\end{tabular}
\end{table*}

\section{Data reduction, Image Analysis and Candidate Detection}
\subsection{Data Reduction}  
The data were reduced using the LuckyCam pipeline.  Standard bias
correction, gain calibration, and cosmic ray removal was applied. The
LuckyCam pipeline registers the image motion of each exposure using an
interpolated cross-correlation algorithm
\citep{Law2006,Staley2010a}. The peak of the cross-correlation map
provides a proxy for the Strehl ratio (i.e. the peak value of the PSF
divided by the theoretical diffraction-limited value, commonly used as
a high-resolution imaging performance metric) and estimates the
relative exposure quality \citep{Staley2010a}.  For each data set,
re-centred and drizzled \citep{Fruchter2002} images are produced by
the pipeline which then selects and co-adds observed frames that meet
the image quality criteria as described in detail by
\citet{Law2006,Staley2010a}. This procedure yields two images for each
data set, the first obtained by co-adding the sharpest 5\%
selection of the frames, and the second by co-adding all exposures
(100\%; see for example Figure~\ref{fig1}--{\em middle panel}). When
choosing the selection cutoff there is a trade-off to be made between
a smaller FWHM at low percentage cutoffs (from fewer images with
higher Strehl ratio), and lower pixel noise at high percentage
cutoffs, due to longer cumulative exposure time.  Figure~\ref{fig1}
shows the improvement obtained with Lucky Imaging for the case of the
planet hosting star, TrES-2 (see also \citealt{Law2006}, Fig.2).

The NOT telescope is subject to aberrations and does not yield near
diffraction-limited images (see
e.g. http://www.not.iac.es/telescope/tti/imqual.pdf). A combination of
small-scale mirror irregularities and chromatic dispersion effects
limit the probability of obtaining diffraction-limited images although
the large number of images and the random phase variation of the
atmosphere can compensate for slight aberrations and telescope
focusing.

Additionally, our Lucky Imaging data do not show ``quasi-static
speckles'', as in adaptive optics imaging (see
e.g. \citealt{Marois2003}; \citealt{Boccaletti2003,Boccaletti2004};
\citealt{Hinkley2007}), that could be mistaken for faint
companions. Our data were visually inspected in order to confirm the
presence of faint companion candidates throughout the data reduction
process. Furthermore, other possible causes of false detection such as
``ghosting'' were not observed in our data.

\begin{figure}
\centering

  \psfig{file=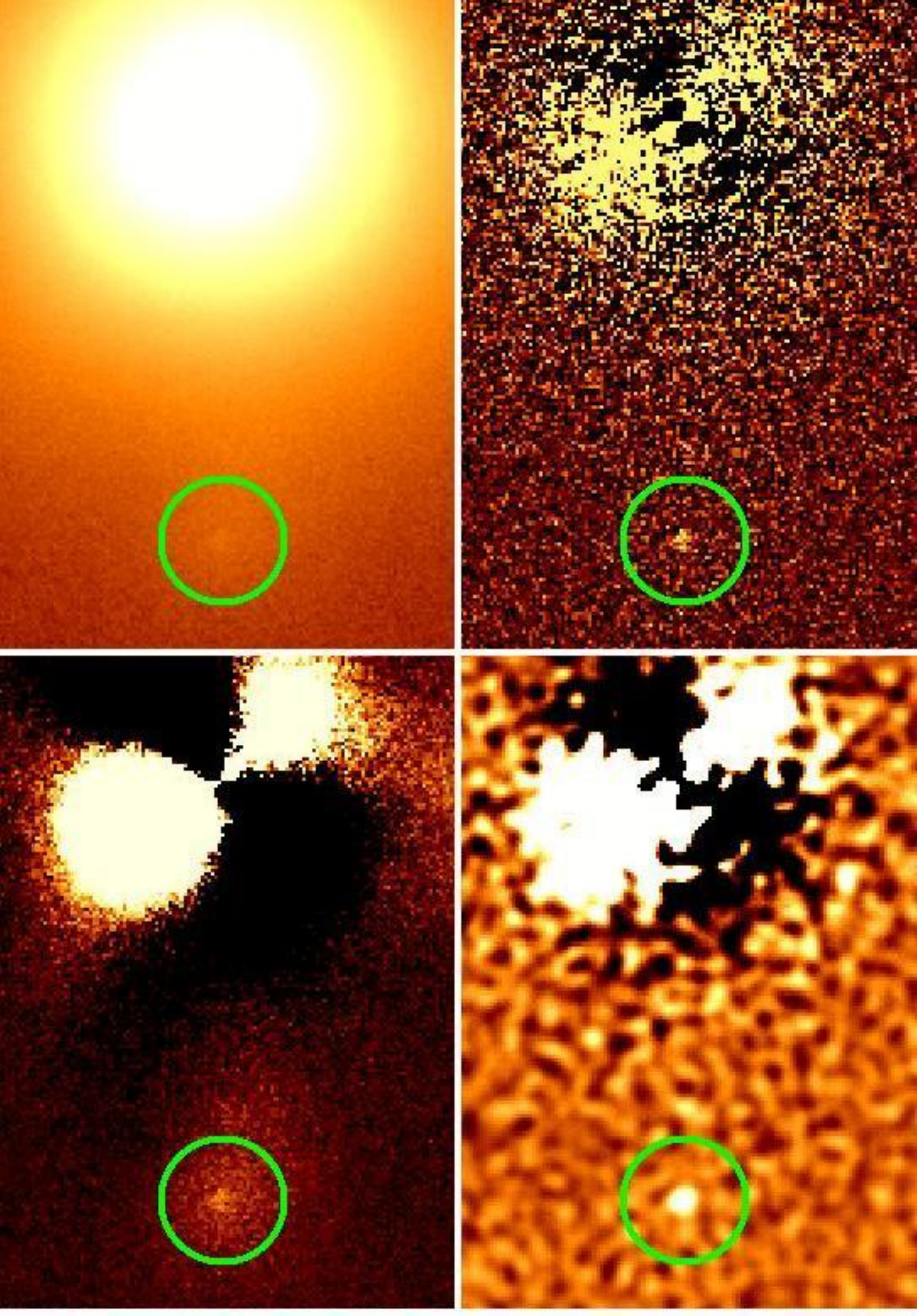,width=0.45\textwidth} 
  \caption{Different stages of the analysis of HAT-P-7. Top-left: The
    image output from the LuckyCam pipeline with the 5\% selection
    criterion.  Bottom-left: After PSF subtraction, step 1 of the
    image analysis algorithm (described in section \S~\ref{analysis}).
    Top-right: After background subtraction via a median boxcar filter
    (Step 2).  Bottom-right: After Gaussian convolution (Step 3).  The
    identified close companion to HAT-P-7 is circled (North is left
    and East is down.)}
\label{figanalysis} 
\end{figure}

\subsection{Image Analysis}
\label{analysis}
The technique most widely applied when attempting to identify faint or
crowded point sources in astronomical images is that of PSF fitting
and subtraction. A step crucial to this process is the choice and
evaluation of PSF models, which may be derived semi-analytically
\citep{Dolphin2000}, empirically \citep{Diolaiti2000}, or by some
combined analytical model fit with empirical corrections
\citep{Stetson1987}. In the case of Lucky Imaging we expect the PSF to
be symmetric, however it is not trivial to model the radial profile as the PSF
consists of a narrow core surrounded by a wide halo \citep[e.g.,][]{Hardy1998}. 
Our image analysis algorithm is described below in three steps:

\begin{enumerate}
\renewcommand{\theenumi}{(\arabic{enumi})}
\item {\bf PSF Subtraction: } To create an axisymmetric,
  semi-empirical model of the PSF we perform a Gaussian fit of 9
  pixels around the brightest, central pixel giving a PSF central
  position to sub-pixel precision. The flux values in the pixels
  around the nominal centre are collected into bins (in radius) and a
  median and standard deviation are evaluated at approximately one
  pixel-width radius intervals. Any visually identified candidate in
  our images is masked off during this process so as not to
  contaminate the PSF model. The Gaussian fit is used within 1.5
  pixels radius from the PSF centre, while at larger radii the model
  is generated using interpolated median values from the annulus
  bins. Finally, the PSF model is subtracted from the original image
  to give a residual image shown
  in Fig.~\ref{figanalysis} (bottom-left).\\*
\item {\bf Background Subtraction via Median Boxcar Filter: } After
  the axisymmetric PSF model has been subtracted, some artefacts can
  remain in the image that might hamper attempts to identify companion
  stars.  In order to validate our detections, we employed a median
  boxcar filter to suppress any artefacts present.  For every pixel,
  the background level is estimated by taking the median of all pixel
  values within a circular aperture of radius 7 pixels (i.e. small
  enough to suppress localised background variations, whilst remaining
  significantly larger than the PSF core so that companion candidates
  are not removed).  The `background map' of median values is then
  subtracted from the residual image (see top-right,
  Fig.~\ref{figanalysis}). \\*

\item {\bf Convolution with a Gaussian Profile: } For this relatively
  small dataset we visually inspected all the sources, utilising a
  Gaussian convolution of the resulting images from Step 2 to enhance
  visibility of any companion candidate (see bottom-right,
  Fig.~\ref{figanalysis}).  Once a candidate has been re-identified,
  the location is inspected in images from all stages of the image
  analysis process (i.e. reduced image, psf-subtracted image, and
  background subtracted image) in order to verify that the candidate
  is not a detector artefact or arising from the image analysis
  process.
\end{enumerate}

\subsection{Candidate Detection}

\begin{figure}
\hspace{-0.6cm}
  \psfig{file=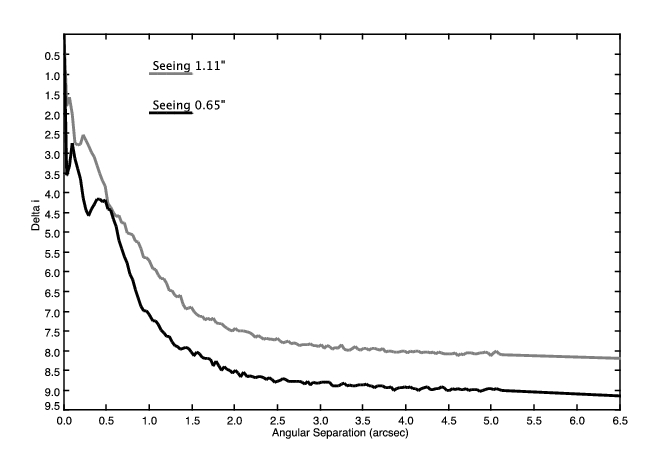,width=0.53\textwidth, height=0.25\textheight}
  \caption[WASP-3]{The effect of seeing conditions on our detection
    sensitivity for the planet host star WASP-3.  The first 10,000
    exposures were obtained with an average seeing of 1.11\arcsec\,
    and the second 5,000 frames with an average seeing of
    0.65\arcsec. The data set obtained during better seeing conditions
    shows an increase in detection sensitivity, important at small
    separations within the seeing disc of the primary star. }
\label{W3} 
\end{figure}

\begin{figure}
  \hspace{-0.4cm}
  \psfig{file=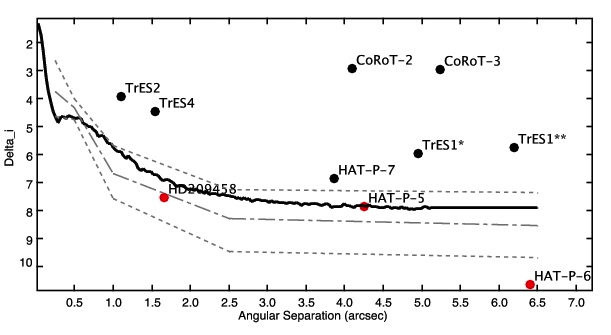,width=0.5\textwidth, height=0.23\textheight} 
  \caption{Average sensitivity of the LuckyCam survey (black line). We
    indicate our detections with black circles and the three
    non-detections discussed in Appendix A with red circles. The grey
    dashed and the dot-dashed lines indicate our minimum, maximum and
    average detection limits (see Table~\ref{limits}) for the sample
    of host stars with no visual companion detected.}
\label{sensitivity} 
\end{figure}

Our detection threshold was chosen to be 4 times the standard
deviation of the background ($\sigma$) at any given concentric circle
at increasing separations from the centre of the planet host.  The
sensitivity of our observations to detect stellar companions at
different angular separations from the primary planet host is given in
Table~\ref{limits}.  We place upper limits in $\Delta i'$ to the
presence of stellar companions to all targets at angular separations
of $0.25\arcsec, 0.5\arcsec, 1\arcsec, 2.5\arcsec$, and $6.5\arcsec$
from the centre of the primary. The adopted 4$\sigma$--detection
limits depend on the exposure time and primary target magnitude as
well as seeing.  This is exemplified in Figure~\ref{W3} where we plot
our detection sensitivity as a function of angular separation in the
case of WASP-3 during observations obtained over two consecutive
nights with different seeing conditions.  The first set of 10,000
images were obtained with an average seeing of 1.11\arcsec\ while the
second 5,000 images were obtained with an average seeing of
0.65\arcsec. The effect of poorer image quality is particularly
evident at small separations within the seeing disc of the planet host
star. Even though, the first set of data have twice the number of
frames, the images taken during better seeing conditions allow the
detection of companions $\Delta i'=$1.8 magnitudes fainter at a
separation of 0.25\arcsec.  Figure~\ref{sensitivity} shows our average
sensitivity. We depict our results in black circles and our
non-detections in red circles. These are discussed in detail in
Appendix A. Additionally we report the minimum, average and maximum
sensitivity curves (grey dashed, dot-dashed lines) derived for the
sample of host stars with no visually detected companions.  Typically
we can detect companions that are $\Delta i'$ $\sim 4$ magnitudes
fainter than the primary at a distance of 0.25\arcsec. As expected,
our sensitivity to fainter companions increases with increasing
distance from the planet host.

Once a candidate companion has been identified, it is verified as a
bona fide stellar source by excluding it as a product of the data
and/or image analysis as follows. First, the FWHM of the planet host
is measured from our images.  Second, the flux of the primary star is
measured using a circular aperture of diameter 6$\times$FWHM on these
images.  Third, on the PSF subtracted images, a Gaussian fit is used
to determine the central pixel position of the candidate companion.
Then, the flux of the identified companion is measured on the PSF
subtracted image, similarly to the primary flux measurement.  Finally,
the signal-to-noise ratio (SNR) of the companion (see
Table~\ref{tab:res}) is calculated taking into account background and
photon shot noise using the following equation:
\begin{equation}
SNR = \frac{F - N_{pix}b}{\sqrt{N_{pix}\sigma_{a}^2+F}}
\label{eq1}
\end{equation}
where $b$ and $\sigma_{a}^{2}$ are the mean value and variance of the
background pixels within the aperture of the companion, $F$ is the
flux over the number of pixels in the photometric aperture,
$N_{pix}$. The SNR values for the candidate companions identified in
this work are given in Table~\ref{tab:res}.

\section{Results}
In the sample of 16 transiting planet host stars we have detected
candidate companion stars for six planet hosts TrES-1, TrES-2, TrES-4,
HAT-P-7, CoRoT-2 and CoRoT-3.  Each candidate companion has been
identified from visual inspection of the reduced Lucky Imaging frames
as described in section \S~4.3.  We summarise our results in
Table~\ref{tab:res} where we give the relative photometry and
astrometry of the companion candidates.  To have a uniform dataset we
only list detections within 6.5\arcsec\ from the centre of the planet
host star.

\begin{table}
  \caption{The 4$\sigma$-detection limits (in $\Delta i'$) for all stars in our sample with and without
    detected companions at separations of $r = 0.25\arcsec,
    0.5\arcsec, 1\arcsec, 2.5\arcsec$, and $6.5\arcsec$. 
  }
\label{limits}
\begin{tabular}{lclllll}
\hline \hline
&\multicolumn{6}{c}{\ \ \ \ \ \ \ 4$\sigma$--detection limits ($\Delta  i'$)} \\ \cline{3-7}
\noalign{\smallskip}
Target & $r(\arcsec)$ & 0.25& 0.5& 1.0 &2.5& 6.5   \\
\hline
HAT-P-1  && 3.57 & 4.12 & 7.62 & 8.99 & 9.22 \\
HAT-P-2  && 4.17 & 4.51 & 6.21 & 8.16 & 8.23 \\
HAT-P-5  && 3.81 & 4.05 & 6.04 & 7.50 & 7.70 \\
HAT-P-6  && 3.63 & 4.78 & 7.50 & 8.87 & 9.12 \\
HAT-P-8  && 4.67 & 4.73 & 5.75 & 7.52 & 7.97 \\
HAT-P-11  && 3.29 & 4.54 & 7.58 & 9.51 & 9.72 \\
HD\,209458  && 4.32 & 4.42 & 7.12 & 9.24 & 9.31 \\
WASP-10  && 3.86 & 4.38 & 6.39 & 7.31 & 7.41 \\
WASP-3$^b$  && 4.36 & 4.03 & 7.08 & 8.78 & 9.18 \\
WASP-3$^a$ & & 2.66 & 4.31 & 5.73 & 7.71 & 8.16 \\
XO-1  && 3.39 & 4.03 & 6.80 & 8.05 & 8.19 \\
\hline
\multicolumn{6}{l}{Targets with candidate companions} \\
\hline
CoRoT-2  && 4.43 & 4.71 & 5.36 & 6.26 & 6.37 \\
CoRoT-3  && 4.90 & 5.09 & 5.51 & 5.92 & 5.89 \\
HAT-P-7  && 4.14 & 4.51 & 5.55 & 7.43 & 7.91 \\
TrES-1  && 4.11 & 4.52 & 7.03 & 7.97 & 8.02 \\
TrES-2  && 4.40 & 5.14 & 6.82 & 7.50 & 7.50 \\
TrES-4  && 4.19 & 4.26 & 6.15 & 7.86 & 7.99 \\
\hline
\multicolumn{6}{l}{$^{a}$ Derived from 10,000, compared to $^b$ 5,000.} \\
\end{tabular}
\end{table}

\begin{figure}
\centering
  \psfig{file=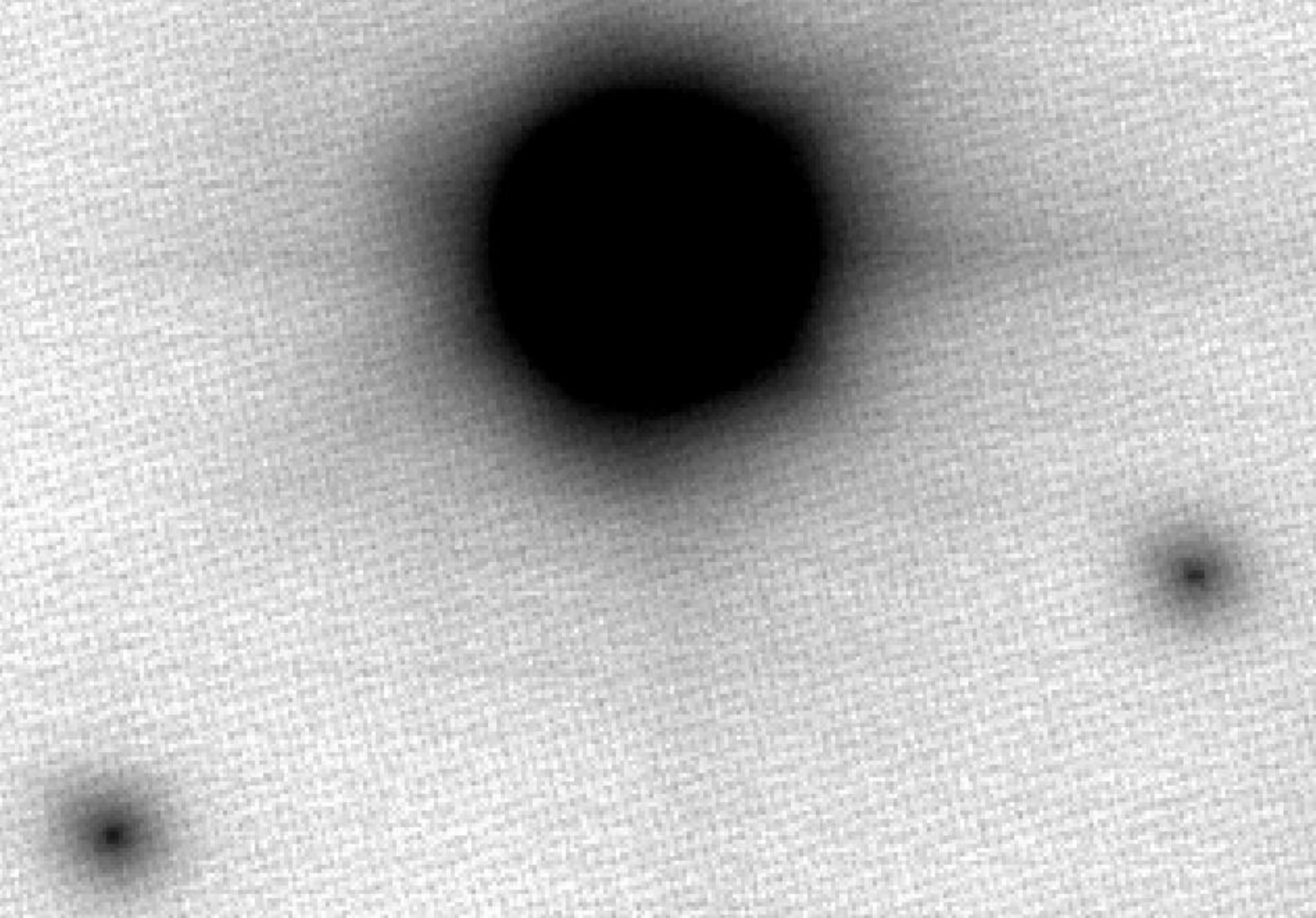,width=0.5\textwidth}
  \caption[TrES-1]{The LuckyCam images for the planet host star
    TrES-1. North is left and East is down. The two companions are
    clearly visible in our images. }
\label{tres1} 
\end{figure}

Our LuckyCam images clearly show the presence of two candidate
companions to the planet hosts star TrES-1, previously
unknown. Figures~\ref{tres1} shows the LuckyCam images for TrES-1 and
the candidate companions identified in this work.

Among our sample TrES-2, TrES-4 and HAT-P-7 have previously published
high-resolution Adaptive Optics (AO) and/or Lucky Imaging observations
showing the presence of faint stellar companions
\citep{Daemgen2009,Narita2010,Bergfors2013}.  Additionally, the
companion stars to CoRoT-3 (2MASS J19281330+0007135) and CoRoT-2
(2MASS J19270636+0122577), have been identified in previous works see
e.g. \citet{Deleuil2008} and \citet{Alonso2008}, \citet{Gillon2010},
respectively. We note that \citet{Deleuil2008} also mentions a second
fainter companion to CoRoT-3 at separation 5.6\arcsec~. We do not
report this object in our discussion as it falls near the CCD edge
making our image analysis unreliable.  The companion to CoRoT-2 and
the two companions to HAT-P-7 have also been confirmed to be bound to
the planet-hosting stars, forming wide binary systems
\citep{Schroter2011,Narita2012}.  We confirm previous findings for the
companions to TrES-2 and TrES-4, while for HAT-P-7 we can only detect
the brighter of the two companions found by
\citet{Narita2010,Narita2012}. The authors estimated the fainter
companion to HAT-P-7 to be of spectral type ${\rm M9-L0}$ (m$_2\simeq
0.078-0.088$ \msun) at a separation of $3.14\pm0.01\arcsec$, and
therefore is below our detection limit for these observations, see
Tables~\ref{limits} and \ref{upperlimits}.  Our results for the
position angles, spectral type determinations, and separations for the
companions to CoRoT-2, TrES-2, TrES-4 and HAT-P-7 agree with the
results obtained by \citet{Alonso2008,Gillon2010,Schroter2011},
\citet{Daemgen2009}, \citet{Narita2010,Narita2012}, and
\citet{Bergfors2013}, and are presented in Tables~\ref{tab:comp} and
\ref{binary}.  In the case of the planet host star HAT-P-8 we are
unable to confirm the candidate companion identified by
\citet{Bergfors2013}. The sensitivity of our observations of HAT-P-8
at the separation of 1.027\arcsec\ would only allow us to detect
companions two magnitudes brighter than the detection reported by the
authors.

\subsection{Non-Detections}
\label{nondetections}

Our visual inspection of the LuckyCam images showed no stellar
companions to the following planet host stars: HAT-P-1, HAT-P-2,
HAT-P-5, HAT-P-6, HAT-P-8, HAT-P-11, HD 209458, WASP-10, WASP-3, and
XO-1. Our results are in agreement with previous studies with the
exception of HAT-P-8 for which our reduced image quality does not
allow the identification of the companion reported by
\citet{Bergfors2013}.  Finally, in the cases of HD 209458, HAT-P-5,
and HAT-P-6 our visual inspection of the images shows possible
candidate companions to the planet hosts, however, after further
consideration (discussed in appendix A), these putative
identifications are classified as non-detections.

\begin{table} 
  \caption{Results for the planet hosting stars with detected 
    candidate stellar companions from this work. From left to right we list the name of 
    the planet host, the angular separation of the candidate companion, the position angle, the $ \Delta
    i'$ magnitude, and the SNR of the detected companion.}
\label{tab:res}
\begin{tabular}{lcccr}
\hline
\hline 
Target 		& $r$  			&PA                    		&$\Delta  i'$		&SNR	\\
                            &(\arcsec)			&($\degree$)		&(mag)              		&	  	\\
\hline  
CoRoT-2		&$4.10\pm0.03$	&$208.4\pm0.4$	&$2.95\pm0.03$	& 41		\\
CoRoT-3		&$5.24\pm0.03$	&$173.9\pm0.4$	&$3.0\pm0.1$		& 10		\\
HAT-P-7		&$3.87\pm0.03$	&$90.4\pm0.5$		&$6.9\pm0.1$		& 10		\\
TrES-1		&$4.95\pm0.03$	&$149.6\pm0.5$	&$6.02\pm0.08$	& 14		\\
TrES-1		&$6.19\pm0.03$	&$ 47.4\pm0.2$	&$5.79\pm0.07$	& 17		\\
TrES-2		&$1.11\pm0.03$	&$137\pm2$		&$3.97\pm0.01$	& 86		\\
TrES-4		&$1.54\pm0.03$	&$1.2\pm1.2$		&$4.51\pm0.02$	& 52		\\
\hline 
\end{tabular}
\end{table}

\section{Statistical Likelihood of Association}
The detection of faint stellar companions associated with our targets
could provide important observational constraints for theoretical
models of planet formation and evolution. We used a statistical
approach to investigate the probability of each detected companion
star being gravitationally bound to the planet host. We first
estimated the density of background sources $\rho(m)$ in a cone of
10\arcmin~around each target. Because our targets are quite bright we
used the 2MASS catalogue to retrieve objects within the 10\arcmin~cone
around the planet host coordinates.  Subsequently, we derive the
probability that a target star has a non-related background source
within the separation of the detected candidate companions.  By using
a similar method to that adopted by \citet{Daemgen2009}, we used the
2MASS magnitudes to identify bright giant stars in the ensemble of
retrieved objects. We selected all objects with ${\rm J-Ks}>0.5$ and
with ${\rm K}<15$, which corresponds to the background detection limit
in these short accumulated exposures (see T$_{\rm exp}$ in
Table~\ref{tab:obs})\footnote{We note however, that for bright guide
  stars longer observations would have allowed the detection of
  background sources as faint as ${\rm i'}\sim22$ (see
  e.g. \citealt{Law2006})}. There is a degeneracy in the near-IR
colours of giant and dwarf stars for early spectral types (earlier
than ${\rm K} 7$ or ${\rm J-Ks}>0.5$), but these become distinct in
two-colour diagrams for the latest spectral types
\citep{Majewski2003}. \citet{Juric2008} used a model of our Galaxy to
estimate the number of giant stars which could be misidentified as
main-sequence stars and found that the overall bias in the estimated
number density is $\sim4\%$ within 500 pc. Finally, we used Equation~1
from \citet{Brandner2000} to find the probability $\mathcal
P(\Theta,m)$ for an unrelated source to be located within a certain
angular distance $\Theta$ from the target.

\begin{equation}
\mathcal P(\Theta,m) = 1-e^{-\pi\rho(m)\Theta^{2}}
\label{giantprob}
\end{equation}
\noindent
where $\Theta$ is in arcsec and $\rho(m)$ is the estimated density of
background sources within 10\arcmin~of the target. We calculated
$\mathcal P(\Theta,m)$ for each star with a detected faint companion
candidate. We also used our images to estimate the expected number of
sources in our images with background, not associated, companions (see
column 6 of Table~\ref{tab:comp}). We note that all but the CoRoT
targets have a very low probability of contamination by background
sources (see Table~\ref{tab:comp}). The CoRoT satellite observes
alternatively towards the galactic centre and anti-centre, thus
increasing the probability of contamination by background objects.

To further test the probability of chance alignment for the binary
pairs we used an independent statistical analysis following the method
described in \citep{Dhital2010}. We calculated the frequency of
unrelated pairings using a Galactic model that is parameterised by an
empirically measured stellar number density distribution in a
$30\arcmin \times 30\arcmin$ conical volume centred on the candidate
binary. The simulated stellar distributions are constrained by
empirical measurements from the Sloan Digital Sky Survey
\citep{Juric2008,Bochanski2010} and accurately accounts for the
decrease in stellar number density with both galactocentric radius and
galactic height. All the simulated stars are, by definition, single
and unrelated. Therefore, the total the number of simulated stars that
are nearby to the candidate primary is the likelihood that the
candidate binary is a chance alignment. We performed 10$^6$
realisations for each of our six candidate binaries.  Table
\ref{tab:comp}, column 6 and 8, show both estimated probabilities
$\mathcal P(\Theta,m)$ and $\mathcal{P}_{D10}$, respectively. Our
results strongly suggest that all the detected faint companions within
6.5\arcsec~to our targets are not random chance alignments.

\subsection{Companion Properties}
Under the assumption that the detected companions are bound to the
planet host stars in our sample, we used 2MASS magnitudes, spectral
types, and temperatures (\teff) of the planet-host targets to derive
spectral types and masses for each candidate companion discussed in
Section \S 5. We first estimated absolute M$_{\rm J}$, M$_{\rm H}$,
M$_{\rm K}$ magnitudes for each planet-host star using their published
distances and 2MASS magnitudes.  We then used the models given in
Table~5 of \citet{Kraus2007}, and models from \citet{Baraffe1998} to
evaluate the absolute $i^{'}$ magnitude for the planet hosts
interpolating within M$_{\rm J}$, M$_{\rm H}$, M$_{\rm K}$, and \teff.
In Table~\ref{tab:comp} we give the estimated M$_{i'}$, spectral types
and masses for each candidate companion. We note that the candidate
companions identified in this study have spectral types later than K4
(see also \citealt{Daemgen2009,Bergfors2013,Narita2010,Schroter2011}),
making them difficult to identify in optical spectra, as well as in
optical, seeing-limited photometry.

The faint stellar companions identified to TrES-1 have separations
from it larger than 2\arcsec, sufficient to avoid blending effects
during spectroscopic and photometric observations.  Such effects in
the case of TrES-2, TrES-4 and HAT-P-7 have been investigated by
\citet{Daemgen2009} and \citet{Bergfors2013} and have been found to be
not significant.  Under the assumptions above, we derived physical
separations, spectral types and masses for the companions to TrES-2,
TrES-4 and HAT-P-7 that are in agreement with previous results (see
Tables~\ref{tab:comp} and \ref{binary}).  For the companion to CoRoT-2
the 2MASS magnitudes are J $= 12.866\pm0.033$, H $=12.234\pm0.044$,
and K $=12.028\pm0.031$. Using the published distance of CoRoT-2 and
the models from \citet{Kraus2007} we obtain a spectral type of M0
($\pm$1 SpT), in agreement with the estimate by \citet{Schroter2011}.
 
The candidate companion to CoRoT-3 is also visible in 2MASS images
\citealt{Cutri2003}, and both stars are classified as 2MASS
J19281330+0007135 and 2MASS 19281326+0007185, respectively. The
near-IR magnitudes of CoRoT-3 are J = $14.027\pm0.036$, H =
$13.448\pm0.045$, and K = $13.295\pm0.043$. The separation between the
objects given in the 2MASS catalogue is $5.1 \pm0.1\arcsec$, in
position angle 173\degree, which are in good agreement with the value
of $5.24\pm0.03\arcsec$ obtained in this work. Our chance alignment
probability for CoRoT-3 is the highest amongst the values derived in
this work, however, the proper motions from the NOMAD catalogue
\citep{Zacharias2005} for CoRoT-3 are $\mu_{\alpha} = -10.7 \pm
5.6$mas/yr and $\mu_{\delta}=21.8$mas/yr, which over the 9 yr span
between the 2MASS and our observations give a total proper motion of
about 0.2\arcsec. Therefore, our results are consistent with the
candidate companion being bound to CoRoT-3.  Assuming the object is at
the same distance as CoRoT-3 we derive a spectral type of K4 -- K5
(see Table~\ref{upperlimits}).

\section{Summary}

To date several different hypotheses have been formulated in order to
explain the observed properties of planetary systems. Compared to our
own solar system, gas giant planets have been found with very short
period orbits (P$<10$d) posing the problem and at the same time,
providing evidence of planetary migration (\citealt{Lin1996},
\citealt{Wu2003}, \citealt{Ida2004}, \citealt{Nagasawa2008},
\citealt{Marzari2009}).  The existence of giant planets in highly
eccentric orbits and the measurements of their spin-orbit
(mis)alignments demonstrate that there must be a number of mechanisms
capable of shaping the system orbital configuration. Although evidence
for such mechanisms has been provided (\citealt{Winn2010b}, \citealt{
  Triaud2010}, \citealt{Hebrard2010}, \citealt{Narita2010},
\citealt{Schlaufman2010}), it is not yet clear which specific
mechanisms are more important or act at a particular time to sculpt
the configuration of known planetary systems. Recently
\citet{Albrecht2012} suggested that the Kozai mechanism is responsible
for the migration of the majority, if not all, hot Jupiters, those
mis-aligned as well as those aligned, and that star-planet tidal
interaction plays a central role in shaping exoplanets orbital
configurations. Moreover, \citet{Narita2012} suggest that the presence
of the two bound companion stars to HAT-P-7 can provide an explanation
of the planetary mis-aligned orbit via sequential Kozai migration
\citep{Takeda2008}. Thus, the detection of faint companions to the
planet hosts will provide important observational evidence,
fundamental for the understanding of the formation and evolution of
their planetary systems.

We have investigated the presence of faint stellar companions within
6.5\arcsec\ of 16 host stars of transiting exoplanets by means of the
Lucky Imaging technique. We show that this technique has the
potential to detect faint stellar companions within the seeing disc
($<1\arcsec$) of bright primary stars.

We have identified faint candidate stellar companions to six planet
hosts. Over the range of brightness of the selected planet host stars
in our sample ($3.50< {\rm M}_{i^{'}}<5.47$, i.e. $7.65<{\rm V}<14$)
we give 4-$\sigma$ detection limits for putative companions at
increasing separations of 0.25\arcsec, 0.5\arcsec, 1\arcsec,
2.5\arcsec~and 6.5\arcsec~from the centre of the primary. For the
targets with no detections, we are able to exclude stellar companions
of spectral types between M1 and M8 at separations $>1\arcsec$,
depending on the brightness of the primary and the seeing at which the
object was observed (see Fig.~\ref{sensitivity}).

We have identified two faint candidate companions to the planet host
TrES-1 that have not been previously reported, and our statistical
analysis suggests that these stars could be bound to the planet host.
Assuming that all the candidate companions are bound to the planet
hosting stars, we used the known distances together with models from
\citet{Kraus2007}, and models from \citet{Baraffe1998} to estimate
spectral types and masses. In the case of TrES-1 we find the first
companion at separation $4.95\pm0.03$\arcsec\ to be of spectral type
M5 ($\pm$1SpT) implying a mass of 0.15M$_{\odot}$. The second at
separation of $6.19\pm0.03$\arcsec\ is found to be of spectral type M5
and mass between 0.2 and 0.15M$_{\odot}$.  In the case of CoRoT-3 we
obtain a spectral type of K4--K5 and a stellar mass between 0.75 and
0.7M$_{\odot}$ for the candidate companion.  For TrES-2, TrES-4,
HAT-P-7, and CoRoT-2 we confirm both known candidates as well as bound
companions and our estimated spectral types and masses agree with
those found by \citet{Daemgen2009,Bergfors2013,Narita2010} and
\citet{Schroter2011,Narita2012}. Overall, for our targets the epoch of
observations either coincide with that of previous works
(e.g. \citealt{Bergfors2013,Narita2012}), or only allow a short
temporal separation with respect to archival and published
observations. Given the precision of our astrometry and the relative
proper motions of the target stars this does not allow any robust
conclusion on the binarity of the detected companions. Therefore,
additional high-resolution high-contrast imaging observations are
necessary in order to robustly confirm if the companions observed in
this and previous works are bound the the planet host stars.

Finally, we discuss in an Appendix the cases of HD 209458, HAT-P-5 and
HAT-P-6, for which possible stellar companions were initially visual
identified in our images but subsequently classified as non-detections
after further analysis was carried out.

\onecolumn

\begin{table} \caption{Upper limits for companions' spectral types and
    masses for the 16 planet host stars in our sample at separations
    $r = 0.25\arcsec, 0.5\arcsec, 1\arcsec$, 2.5\arcsec, and
    6.5\arcsec~from the primary.}
\label{upperlimits}
\begin{tabular}{lcccccccccc}
\hline \hline 
&\multicolumn{2}{c}{$r = 0.25\arcsec$}&\multicolumn{2}{c}{$r = 0.5\arcsec$}&\multicolumn{2}{c}{$r = 1\arcsec$} &\multicolumn{2}{c}{$r = 2.5\arcsec$} &\multicolumn{2}{c}{$r = 6.5\arcsec$} \\  \cline{2-11} 
\noalign{\smallskip}
Target 	& Sp.T&  m$_2$ &Sp.T &  m$_2$ & Sp.T & m$_2$& Sp.T&  m$_2$& Sp.T&  m$_2$\\		
&$\pm1$ &\msun&$\pm1$ &\msun&$\pm1$&\msun &$\pm1$&\msun&$\pm1$&\msun \\	
\hline
HAT-P-1		&K7--M0 	&0.63--0.59	&M1--M2 	&0.54--0.42	&M5		&0.15		&M7		&0.11		&M7		&0.11	\\
HAT-P-2		&M0--M1 	&0.59--0.54	&M1--M2	&0.54--0.42	&M4--M5	&0.20--0.15	&M7		&0.11		&M7		&0.11	\\
HAT-P-5		&M0		&0.59		&M1		&0.54		&M4--M5	&0.20--0.15	&M6	 	&0.12		&M6	 	&0.12	\\
HAT-P-6		&M0		&0.59		&M2		&0.42		&M5		&0.15	 	&M6		&0.12		&M7	 	&0.11	\\
HAT-P-8		&M2		&0.42		&M2		&0.42		&M4		&0.20		&M6		&0.12		&M6		&0.12	\\
HAT-P-11	&M2		&0.42		&M5		&0.15		&M6--M7	&0.12--0.11	&L0	 	&0.078		&$>$L0	&$<$0.078	\\
HD\,209458	&M1--M2	&0.54--0.42	&M2		&0.42		&M5		&0.15	 	&M6--M7	&0.12--0.11	&M7  	&0.11	\\
WASP-3$^a$	&K5		        &0.70		&M1		&0.54		&M4--M5 	&0.20--0.15 	&M6		&0.12		&M6		&0.12	\\
WASP-3$^b$	&M0		&0.59		&M0--M1 	&0.59--0.54	&M4		&0.20	 	&M7		&0.11		&M7		&0.11	\\
WASP-10		&M4		&0.20		&M4--M5	&0.20--0.15	&M6--M7	&0.12--0.11	&M8		&0.102		&M8		&0.102	\\
XO-1		&K7                 &0.63		&M1--M2	&0.54--0.42	&M5		&0.15 		&M6--M7	&0.12--0.11	&M6--M7	&0.12--0.11	\\
\hline
\multicolumn{4}{l}{Targets with companion candidates} \\
\hline
CoRoT-2		&M2--M3	&0.42--0.29	&M3		&0.29		&M4		&0.20		&M4--M5	&0.20--0.15	&M5	 	&0.15	\\
CoRoT-3		&M2		&0.42		&M1		&0.54		&M2		&0.42		&M3		&0.29		&M3  	&0.29	\\
HAT-P-7		&M1		&0.54		&M1		&0.54		&M3--M4	&0.29--0.20 	&M5--M6	&0.15--0.12	&M6  	&0.12	\\
TrES-1		&M2--M3	&0.42--0.29	&M3--M4 	&0.29--0.20	&M6		&0.12		&M7--M8	&0.11--0.102	&M7  	&0.11	\\
TrES-2		&M1--M2	&0.54--0.42	&M3		&0.29		&M5		&0.15	 	&M6--M7	&0.12--0.11	&M6--M7  &0.12--0.11	\\
TrES-4		&M0		&0.59		&M1		&0.54		&M4--M5	&0.20--0.15 	&M6		&0.12		&M6--M7	&0.12--0.11	\\
\hline
\end{tabular}
\end{table}

\begin{table} 
  \caption{Companion candidates for 6 planet host stars.
    From left to right we list the name of the planet host star, separation angle, the position angle, the $ \Delta
    i'$ for the detected companions, the SNR of the detected companion, the probability for the companion to be a chance alignment ($\mathcal{P}(\Theta,m)$) and the 
    expected number of sources with an unrelated background companion (E$_{\rm bg}$), the probability of a chance alignment detection as estimated by \citet{Dhital2010}, the planet host's distance (pc), and finally 
    the companion separation in AU, assuming the value is a lower limit.}
  \label{tab:comp}
\begin{tabular}{lcccrllccc}
\hline
\hline 
Target 		& $r$  			&PA                    		&$\Delta  i'$		&SNR&$\mathcal{P}(\Theta,m)$ 	& E$_{\rm bg}$ &$\mathcal{P}_{D10}$	&Dist. &Sep.\\                
&(\arcsec)	&($\degree$)		&(mag)              	&	  &(\%)                  			&			&(\%)				&(pc)  &(AU)  \\
\hline  
CoRoT-2		&$4.10\pm0.03$	&$208.4\pm0.4$	&$2.95\pm0.03$	& 41	&3.17	&0.22	&1.18	&$270\pm120$	&$1108\pm492$\\    
CoRoT-3		&$5.24\pm0.03$	&$173.9\pm0.4$	&$3.0\pm0.1$	& 10	&4.05	&0.12	&1.72	&$680\pm160$	&$3562\pm838$\\  
HAT-P-7		&$3.87\pm0.03$	&$90.4\pm0.5$	&$6.9\pm0.1$	& 10	&0.03	&0.004	&0.2	&$320\pm50$	&$1238\pm193$ \\   
TrES-1		&$4.95\pm0.03$	&$149.6\pm0.5$	&$6.02\pm0.08$	& 14	&0.82	&0.025	&0.04	&$150\pm6$	&$743\pm30$  \\       
TrES-1		&$6.19\pm0.03$	&$ 47.4\pm0.2$	&$5.79\pm0.07$	& 17	&1.29	&0.039	&0.06	&$150\pm6$	&$929\pm37$  \\     
TrES-2		&$1.11\pm0.03$	&$137\pm2$		&$3.97\pm0.01$	& 86	&0.03	&0.0005	&0		&$220\pm10$	&$244\pm13$  \\    
TrES-4		&$1.54\pm0.03$	&$1.2\pm1.2$	&$4.51\pm0.02$	& 52	&0.03	&0.0007	&0		&$479\pm26$	&$740\pm43$  \\   
\hline 
\end{tabular}
\end{table}

\twocolumn

\begin{landscape}
 \centering
\begin{table} 
  \caption{Estimated absolute $i'$ magnitudes (M$_i'$), spectral types and masses for the 
    companion stars, derived assuming binarity for each companion. Values for the companions are derived from \citet{Kraus2007} and \citet{Baraffe1998} models using published 2MASS magnitudes, distances and \teff~of the planet host targets. Magnitude errors are estimated through propagation of the known errors on the target J, H, K magnitude and distances. Superscript $^1$ and $^2$ indicate the host star and the companion(s), respectively.  }
\label{binary}

\begin{tabular}{lcccccccclrrcc}
\hline
\hline
Target &J$^1$&H$^1$&K$^1$&M$^1_{\, \rm J}$&M$^1_{\, \rm H}$&M$^1_{\, \rm K}$&SpT$^1$&T$^{1}_{eff}$&M$^1_{i^{'}}$&${\rm \Delta i'}$&M$^{2}_{i^{'}}$&SpT$^2$&m$_2$ \\ 
&&&&&&&&(K)&&(mag)&&&(\msun)\\ 
\hline     
CoRoT-2$^{\dagger}$        &  $10.78\pm 0.02$  & $10.44\pm 0.02$  & $10.31\pm 0.019$  &       $ 3.63\pm 0.20$  &      $3.28\pm 0.22$ &        $3.15\pm 0.22$ &  G7   &  $ 5608\pm37  $  &  4.60   &  2.95  & 7.55&  M0             &0.59          \\
CoRoT-3        &  $11.94\pm 0.02$  & $11.71\pm 0.02$  & $11.62\pm 0.019$  &       $2.77\pm 0.22 $ &       $2.55\pm 0.22$ &        $2.46\pm 0.22$ &  F3    &  $ 6740\pm140$ &  3.50   &  3.00  & 6.45&  K4--K5     &0.75--0.70\\      
HAT-P-7$^{\dagger}$        &    $9.55\pm 0.02$  &  $9.34\pm 0.02$    & $9.33\pm 0.020  $  &        $2.03\pm 0.29 $ &       $1.82\pm 0.29$ &        $1.81\pm 0.29$ &  F6    &  $ 6350\pm80  $ &  4.00   &  6.92  &10.92&  M4--M5   &0.20--0.15  \\
TrES-1           & $10.29\pm  0.03$  & $9.89\pm 0.04$   & $9.82\pm 0.030  $  &         $4.41\pm 2.25 $ &       $4.01\pm 2.26$ &        $3.94\pm 2.25$ &  K0   &  $ 5214\pm23  $ &  5.47  &   6.02  &11.49&  M5            &0.15               \\
TrES-1           & $10.29\pm  0.03$  & $9.89\pm 0.04$   & $9.82\pm 0.030  $  &         $4.41\pm 1.21 $ &       $4.01\pm 1.22$ &        $3.94\pm 1.21$ &  K0   &  $ 5214\pm23  $ &  5.47  &   5.80  &11.27&  M5            &0.20--0.15     \\
TrES-2           & $10.23\pm  0.03$  & $9.93\pm 0.03$   & $9.85\pm 0.020  $  &         $3.52\pm 0.81 $ &       $3.21\pm 0.81$ &        $3.13\pm 0.80$ &  G0   &  $ 5850\pm50  $ &  4.44  &   3.97  & 8.41&  M1--M2    &0.54--0.42  \\
TrES-4           & $10.58\pm  0.02$  & $10.35\pm0.02$  & $10.33\pm 0.020$  &        $2.18\pm 0.40 $ &       $1.95\pm 0.40$ &        $1.93\pm 0.40$ &  F8    &  $ 6200\pm75  $ &  4.26  &   4.50  & 8.76&  M2            &0.42            \\
\hline
\multicolumn{4}{l}{Non-detections} \\
\hline
HD\,209458  &   $6.59\pm  0.02$  & $6.37\pm 0.02$   & $6.31\pm 0.020  $  &         $3.23\pm 0.46 $ &       $3.00\pm 0.46$ &        $2.94\pm 0.46$ &  G0   &  $ 6075\pm33  $ &  4.40   &  7.57  &11.97&  M5--M6   &0.15--0.12             \\         
HAT-P-5       &  $10.84\pm 0.02$  &  $10.52\pm 0.03$ & $10.48\pm 0.020$  &        $3.17\pm 0.25 $ &       $2.85\pm 0.26$ &        $2.81\pm 0.25$ &  G1   &  $ 5960\pm100$ &  4.40   &  7.91  &12.31&  M6            &0.12          \\
HAT-P-6       &    $9.56\pm 0.02$  &  $ 9.44\pm 0.04$   & $9.31\pm 0.030  $  &        $2.47\pm 0.25 $ &       $2.36\pm 0.27$ &        $2.23\pm 0.26$ &  F8    &  $ 6570\pm80  $ &  3.69   & 10.69 &14.38&  M7--M8   &0.10             \\       
\hline 
\multicolumn{6}{l}{$^{\dagger}$ Confirmed bound companions} \\
\end{tabular}	
\end{table}
\end{landscape}

\section*{ACKNOWLEDGEMENTS}

Based on observations made with the Nordic Optical Telescope, operated
on the island of La Palma jointly by Denmark, Finland, Iceland,
Norway, and Sweden, in the Spanish Observatorio del Roque de los
Muchachos of the Instituto de Astrofisica de Canarias. FF would like
to thank STFC for support through the award of a PDRA as part of the
QUB Rolling Grant for Astrophysics. Y.G.M.C. acknowledges postdoctoral
funding support from the Vanderbilt Office of the Provost, through the
Vanderbilt Initiative in Data-intensive Astrophysics (VIDA) and
through a grant from the Vanderbilt International Office in support of
the Vanderbilt-Warwick Exoplanets Collaboration.

\appendix

\section{Non Detections with visual identifications}
\begin{enumerate}

\item {\bf HD 209458:} For the planet host star HD 209458, slight
  aberration effects are evident in our images resulting from small
  scale mirror irregularities of the NOT, and chromatic dispersion
  effects \citep{Law2006}.  These effects are more pronounced in the
  images of bright targets like HD 209458 \citep[V=7.63;][]{Hog2000}.
  The possible detection was present in all four stages of our image
  analysis at a separation of 1.66\arcsec\ (within the seeing disc of
  the primary star) and position angle $241\pm 1$\degree~with $\Delta
  i'$ = 7.57 (SNR $\sim$20).  Figure~\ref{hd} presents the PSF
  subtraction and the Gaussian convolution steps for HD 209458 showing
  evidence of the non-axisymmetric PSF, and of the possible
  detection. Figure \ref{hd_limits} shows our sensitivity as a
  function of separation from the centre of the primary target.  Our
  possible detection is well above our sensitivity limit at the
  separation of 1.66\arcsec.  However, any identification in our
  images within the seeing disc of the planet host is investigated
  further for possible artefacts.  VLT+NACO images in the $H$-band for
  HD 209458 are publicly available from the ESO archive\footnote{ESO
    Archive: http://archive.eso.org/cms.html}.  Our analysis of these
  NACO near-infrared adaptive optic data do not show any evidence of a
  stellar companion at the position of our possible detection.  We
  would have expected any stellar companion to be brighter in the
  near-infrared, and thus be readily identifiable in the NACO
  photometry.  This is also in agreement with the non detection in the
  Lucky Imaging observations by \citet{Daemgen2009} and
  \citet{Bergfors2013}.  Therefore, we conclude that the possible
  detection is most likely spurious due to the limited image quality
  for HD 209458, resulting from the seeing conditions,
  the number of frames, and the optical characteristics of the NOT.  \\*

\begin{figure}
  \centering
  \psfig{file=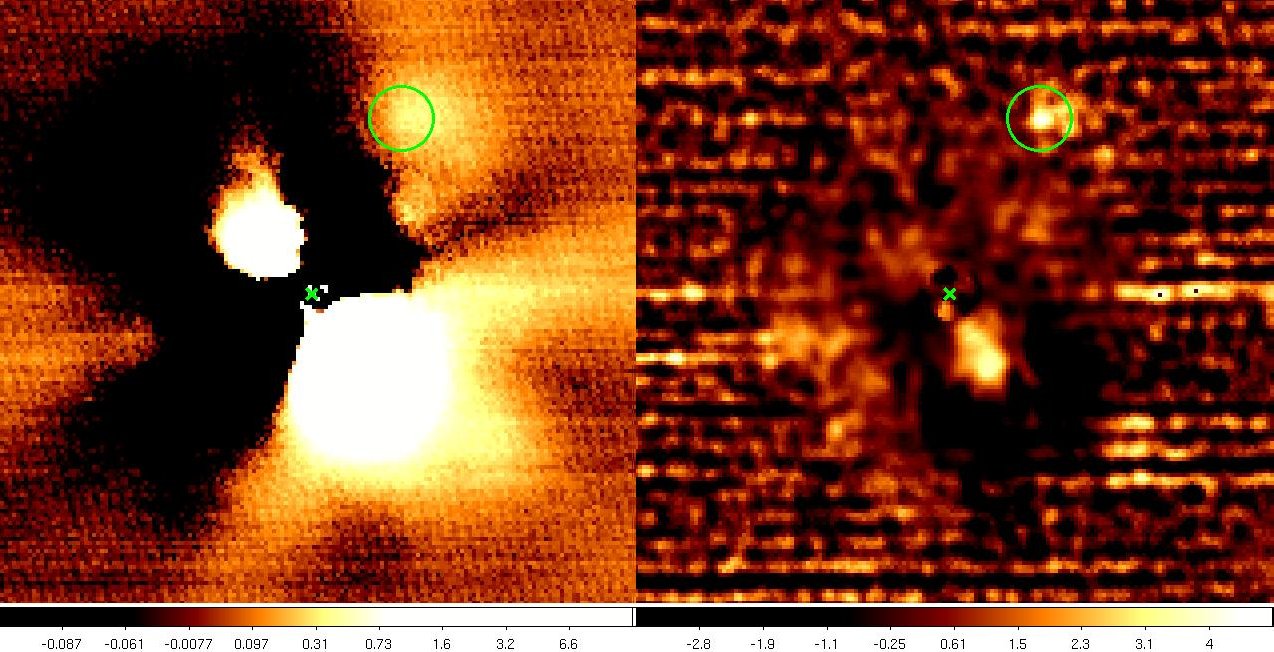,width=0.45\textwidth} 
  \caption[HD209458]{Spurious detection for HD 209458 at a separation
    of 1.66\arcsec\ with a $\Delta i' = 7.57$.  {\it Left panel:}
    Image after PSF subtraction (Step 1 of image analysis). {\it Right
      panel:} Image after Gaussian convolution (Step 3). The green
    cross marks the centre of the primary star, whereas circled in
    green is the spurious detection most likely due to our image
    quality.}
\label{hd} 
\end{figure}

 \begin{figure}
\centering
 \psfig{file=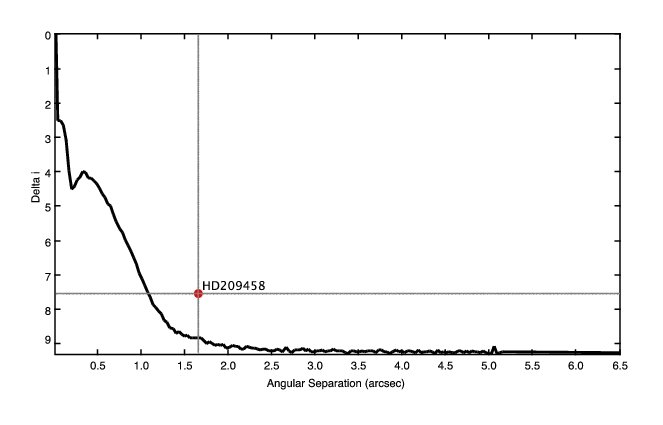,width=0.5\textwidth} 
 \caption[limits]{Sensitivity curve as function of distance from the
   primary plant host star HD 209458 derived for the 5\% best-frame
   selection. The vertical and horizontal grey-solid lines indicate
   the angular separation ($r=1.66\arcsec$), and $\Delta i'=5.57$ of
   the possible detection, respectively.  Our sensitivity at the
   angular separation of 1.66\arcsec\ is $\Delta i'=8.9$. }
\label{hd_limits}
\end{figure}
 
 \begin{figure}
\centering
  \psfig{file=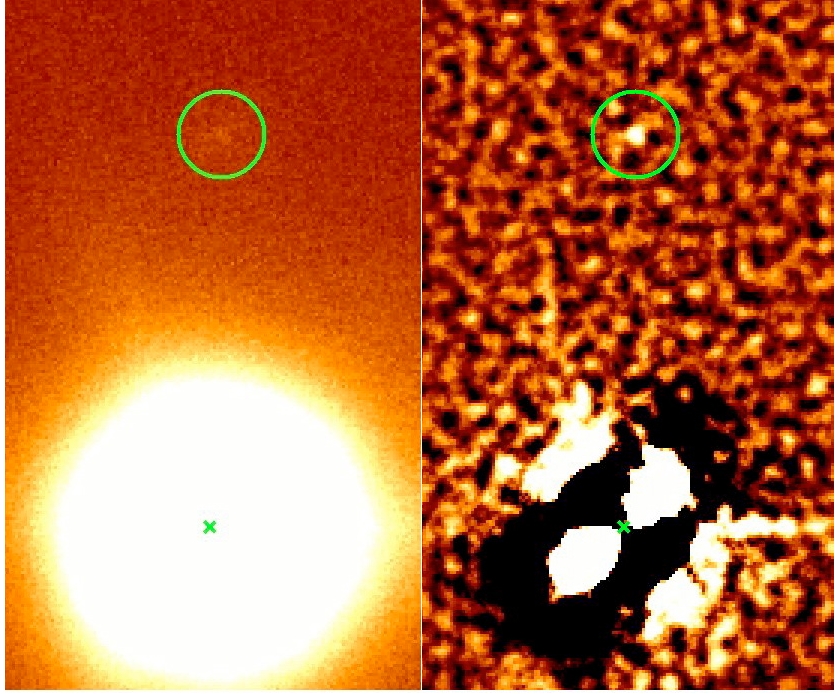,width=0.39\textwidth,height=0.21\textheight} 
  \caption{Non-detection for HAT-P-5. {\it Left panel:} LuckyCam
    5\%-frame selection image for HAT-P-5 (Step 1). {\it Right panel:}
    The Gaussian convolution image (Step 3).  The green cross marks
    the location of the centre of the primary star, the tentative
    companion is circled in green.}
\label{H5} 
\end{figure}

\begin{figure}
  \centering \psfig{file=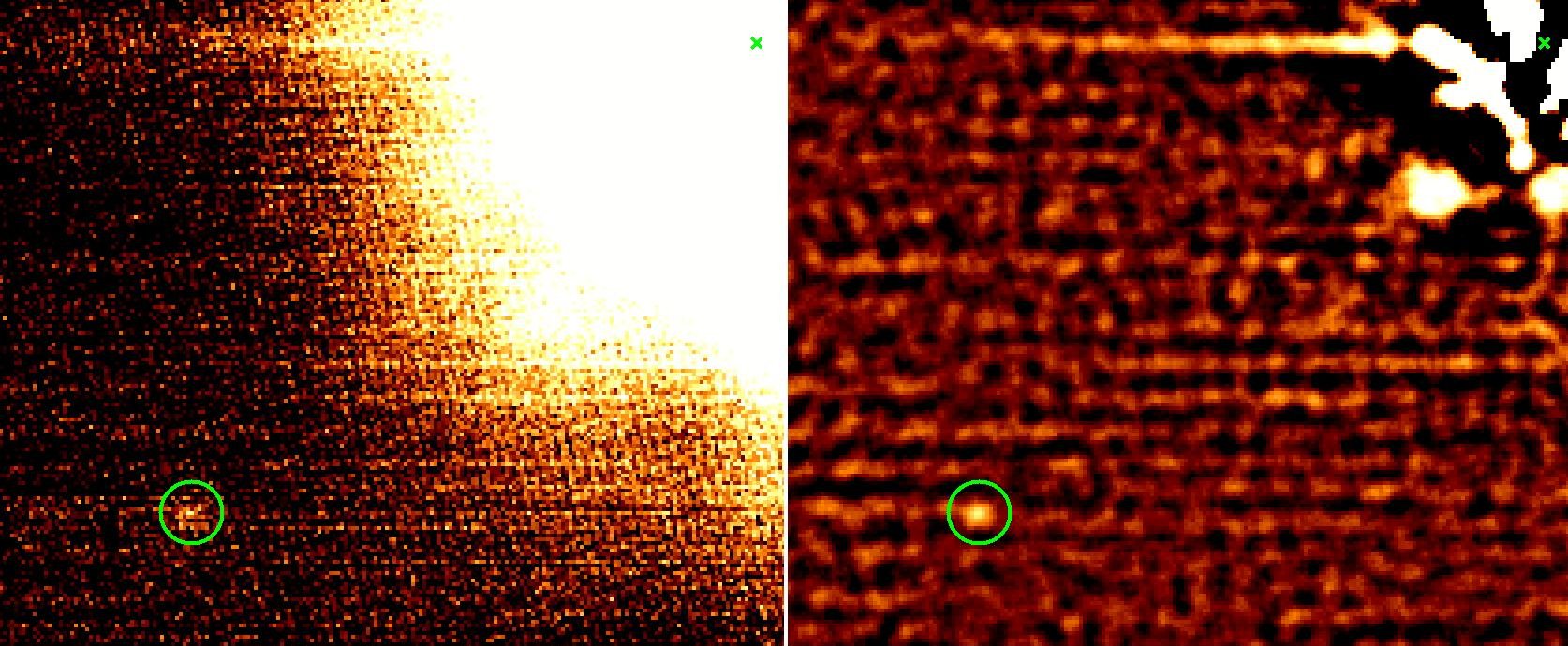,width=0.45\textwidth}
  \caption{Non-detection for HAT-P-6. {\it Left panel:} LuckyCam
    5\%-frame selection image for HAT-P-6 (Step 1). {\it Right panel:}
    The Gaussian convolution image (Step 3).  The green cross marks
    the location of the centre of the primary star, the tentative
    companion is circled in green.}
\label{fig:HAT-P-6} 
\end{figure}

\item {\bf HAT-P-5:} During our image analysis procedure and visual
  inspection of the images for the planet host HAT-P-5 we have
  identified a candidate companion with $\Delta i' = 7.9$
  (SNR$\sim$1.9) at a separation of 4.25\arcsec\ from the centre of
  the primary star and position angle $268.5\pm0.4$\degree.
  Figure~\ref{H5} shows the image from the 5\% best LuckyCam frames
  for HAT-P-5 (left), and the Step 3 (right) of the image analysis
  where the candidate companion is clearly visible.
  The measured $\Delta i'$ is 0.14 magnitude below our 4-$\sigma$
  detection cut-off at the separation of 4.25\arcsec, thus it was
  classified as a non-detection. \\*

\item {\bf HAT-P-6:} In the images of the planet host HAT-P-6, a
  candidate companion with $\Delta i'$ $\approx 10.7$ (corresponding
  to a SNR $\sim$0.4) at a separation of 6.4\arcsec\ is identified by
  visual inspection. For example, Figure~\ref{fig:HAT-P-6} shows the
  LuckyCam images for Step 1 (left) and Step 3 (right) of the image
  analysis where the candidate companion is clearly visible.  However,
  the measured $\Delta i'$ of the putative companion is more than one
  magnitude below the 4-$\sigma$ detection threshold at that
  separation from the centre of the primary (see Table~\ref{limits}),
  thus it is considered a non-detection.\\*

  In the case of HAT-P-5 and HAT-P-6 our image sensitivity does not
  allow us to reliably detect the putative companions.  However,
  because our images clearly show the presence of possible companions
  at large separations from the primary, these might be real and worth
  further investigation.

\end{enumerate}

\bibliographystyle{mn2e}
\bibliography{lucky.bib}

\begin{thebibliography}{}

\bibitem[\protect\citeauthoryear{{Albrecht}, {Winn}, {Johnson}, {Howard},
  {Marcy}, {Butler}, {Arriagada}, {Crane}, {Shectman}, {Thompson}, {Hirano},
  {Bakos} \& {Hartman}}{{Albrecht} et~al.}{2012}]{Albrecht2012}
{Albrecht} S.,  {Winn} J.~N.,  {Johnson} J.~A.,  {Howard} A.~W.,  {Marcy}
  G.~W.,  {Butler} R.~P.,  {Arriagada} P.,  {Crane} J.~D.,  {Shectman} S.~A.,
  {Thompson} I.~B.,  {Hirano} T.,  {Bakos} G.,    {Hartman} J.~D.,  2012, ApJ,
  757, 18

\bibitem[\protect\citeauthoryear{{Alonso}, {Auvergne}, {Baglin}, {Ollivier},
  {Moutou}, {Rouan}, {Deeg} et~al.,}{{Alonso} et~al.}{2008}]{Alonso2008}
{Alonso} R.,  {Auvergne} M.,  {Baglin} A.,  {Ollivier} M.,  {Moutou} C.,
  {Rouan} D.,  {Deeg} H.~J.,    et~al., 2008, A\&A, 482, L21

\bibitem[\protect\citeauthoryear{{Anderson}, {Hellier}, {Gillon}
  et~al.,}{{Anderson} et~al.}{2010}]{Anderson2010}
{Anderson} D.~R.,  {Hellier} C.,  {Gillon} M.,    et~al., 2010, ApJ, 709, 159

\bibitem[\protect\citeauthoryear{{Bakos}, {Howard}, {Noyes} et~al.,}{{Bakos}
  et~al.}{2009}]{Bakos2009}
{Bakos} G.~{\'A}.,  {Howard} A.~W.,  {Noyes} R.~W.,    et~al., 2009, ApJ, 707,
  446

\bibitem[\protect\citeauthoryear{{Bakos}, {Noyes}, {Kov{\'a}cs}
  et~al.,}{{Bakos} et~al.}{2007}]{Bakos2007}
{Bakos} G.~{\'A}.,  {Noyes} R.~W.,  {Kov{\'a}cs} G.,    et~al., 2007, ApJ, 656,
  552

\bibitem[\protect\citeauthoryear{{Baldwin}, {Tubbs}, {Cox}, {Mackay}, {Wilson}
  \& {Andersen}}{{Baldwin} et~al.}{2001}]{Baldwin2001}
{Baldwin} J.~E.,  {Tubbs} R.~N.,  {Cox} G.~C.,  {Mackay} C.~D.,  {Wilson}
  R.~W.,    {Andersen} M.~I.,  2001, A\&A, 368, L1

\bibitem[\protect\citeauthoryear{{Baraffe}, {Chabrier}, {Allard} \&
  {Hauschildt}}{{Baraffe} et~al.}{1998}]{Baraffe1998}
{Baraffe} I.,  {Chabrier} G.,  {Allard} F.,    {Hauschildt} P.~H.,  1998, A\&A,
  337, 403

\bibitem[\protect\citeauthoryear{{Bergfors}, {Brandner}, {Daemgen}, {Biller},
  {Hippler}, {Janson}, {Kudryavtseva}, {Gei{\ss}ler}, {Henning} \&
  {K{\"o}hler}}{{Bergfors} et~al.}{2013}]{Bergfors2013}
{Bergfors} C.,  {Brandner} W.,  {Daemgen} S.,  {Biller} B.,  {Hippler} S.,
  {Janson} M.,  {Kudryavtseva} N.,  {Gei{\ss}ler} K.,  {Henning} T.,
  {K{\"o}hler} R.,  2013, MNRAS, 428, 182

\bibitem[\protect\citeauthoryear{{Boccaletti}, {Chauvin}, {Lagrange} \&
  {Marchis}}{{Boccaletti} et~al.}{2003}]{Boccaletti2003}
{Boccaletti} A.,  {Chauvin} G.,  {Lagrange} A.-M.,    {Marchis} F.,  2003,
  A\&A, 410, 283

\bibitem[\protect\citeauthoryear{{Boccaletti}, {Riaud}, {Baudoz}, {Baudrand},
  {Rouan}, {Gratadour}, {Lacombe} \& {Lagrange}}{{Boccaletti}
  et~al.}{2004}]{Boccaletti2004}
{Boccaletti} A.,  {Riaud} P.,  {Baudoz} P.,  {Baudrand} J.,  {Rouan} D.,
  {Gratadour} D.,  {Lacombe} F.,    {Lagrange} A.-M.,  2004, PASP, 116, 1061

\bibitem[\protect\citeauthoryear{{Bochanski}, {Hawley}, {Covey}, {West},
  {Reid}, {Golimowski} \& {Ivezi{\'c}}}{{Bochanski}
  et~al.}{2010}]{Bochanski2010}
{Bochanski} J.~J.,  {Hawley} S.~L.,  {Covey} K.~R.,  {West} A.~A.,  {Reid}
  I.~N.,  {Golimowski} D.~A.,    {Ivezi{\'c}} {\v Z}.,  2010, AJ, 139, 2679

\bibitem[\protect\citeauthoryear{{Brandner}, {Zinnecker}, {Alcal{\'a}},
  {Allard}, {Covino}, {Frink}, {K{\"o}hler}, {Kunkel}, {Moneti} \&
  {Schweitzer}}{{Brandner} et~al.}{2000}]{Brandner2000}
{Brandner} W.,  {Zinnecker} H.,  {Alcal{\'a}} J.~M.,  {Allard} F.,  {Covino}
  E.,  {Frink} S.,  {K{\"o}hler} R.,  {Kunkel} M.,  {Moneti} A.,
  {Schweitzer} A.,  2000, AJ, 120, 950

\bibitem[\protect\citeauthoryear{{Burrows}, {Hubeny}, {Budaj} \&
  {Hubbard}}{{Burrows} et~al.}{2007}]{Burrows2007}
{Burrows} A.,  {Hubeny} I.,  {Budaj} J.,    {Hubbard} W.~B.,  2007, ApJ, 661,
  502

\bibitem[\protect\citeauthoryear{{Chatterjee}, {Ford}, {Matsumura} \&
  {Rasio}}{{Chatterjee} et~al.}{2008}]{Chatterjee2008}
{Chatterjee} S.,  {Ford} E.~B.,  {Matsumura} S.,    {Rasio} F.~A.,  2008, ApJ,
  686, 580

\bibitem[\protect\citeauthoryear{{Crida}, {S{\'a}ndor} \& {Kley}}{{Crida}
  et~al.}{2008}]{Crida2008}
{Crida} A.,  {S{\'a}ndor} Z.,    {Kley} W.,  2008, A\&A, 483, 325

\bibitem[\protect\citeauthoryear{{Crossfield}, {Barman}, {Hansen}, {Tanaka} \&
  {Kodama}}{{Crossfield} et~al.}{2012}]{Crossfield2012}
{Crossfield} I.~J.~M.,  {Barman} T.,  {Hansen} B.~M.~S.,  {Tanaka} I.,
  {Kodama} T.,  2012, ApJ, 760, 140

\bibitem[\protect\citeauthoryear{{Cutri}, {Skrutskie}, {van Dyk}
  et~al.,}{{Cutri} et~al.}{2003}]{Cutri2003}
{Cutri} R.~M.,  {Skrutskie} M.~F.,  {van Dyk} S.,    et~al., 2003, VizieR
  Online Data Catalog, 2246, 0

\bibitem[\protect\citeauthoryear{{Daemgen}, {Hormuth}, {Brandner}, {Bergfors},
  {Janson}, {Hippler} \& {Henning}}{{Daemgen} et~al.}{2009}]{Daemgen2009}
{Daemgen} S.,  {Hormuth} F.,  {Brandner} W.,  {Bergfors} C.,  {Janson} M.,
  {Hippler} S.,    {Henning} T.,  2009, A\&A, 498, 567

\bibitem[\protect\citeauthoryear{{Deleuil}, {Deeg}, {Alonso}, {Bouchy},
  {Rouan}, {Auvergne}, {Baglin} et~al.,}{{Deleuil} et~al.}{2008}]{Deleuil2008}
{Deleuil} M.,  {Deeg} H.~J.,  {Alonso} R.,  {Bouchy} F.,  {Rouan} D.,
  {Auvergne} M.,  {Baglin} A.,    et~al., 2008, aap, 491, 889

\bibitem[\protect\citeauthoryear{{Dhital}, {West}, {Stassun} \&
  {Bochanski}}{{Dhital} et~al.}{2010}]{Dhital2010}
{Dhital} S.,  {West} A.~A.,  {Stassun} K.~G.,    {Bochanski} J.~J.,  2010, AJ,
  139, 2566

\bibitem[\protect\citeauthoryear{Diolaiti, Bendinelli, Bonaccini, Close, Currie
  \& Parmeggiani}{Diolaiti et~al.}{2000}]{Diolaiti2000}
Diolaiti E.,  Bendinelli O.,  Bonaccini D.,  Close L.,  Currie D.,
  Parmeggiani G.,  2000, A\&AS, 147, 335

\bibitem[\protect\citeauthoryear{Dolphin}{Dolphin}{2000}]{Dolphin2000}
Dolphin A.~E.,  2000, PASP, 112, 1383

\bibitem[\protect\citeauthoryear{{Eggenberger}, {Udry} \&
  {Mayor}}{{Eggenberger} et~al.}{2004}]{Eggenberger2004}
{Eggenberger} A.,  {Udry} S.,    {Mayor} M.,  2004, A\&A, 417, 353

\bibitem[\protect\citeauthoryear{{Fabrycky} \& {Tremaine}}{{Fabrycky} \&
  {Tremaine}}{2007}]{Fabrycky2007}
{Fabrycky} D.,  {Tremaine} S.,  2007, ApJ, 669, 1298

\bibitem[\protect\citeauthoryear{{Fortney}, {Marley} \& {Barnes}}{{Fortney}
  et~al.}{2007}]{Fortney2007}
{Fortney} J.~J.,  {Marley} M.~S.,    {Barnes} J.~W.,  2007, ApJ, 659, 1661

\bibitem[\protect\citeauthoryear{{Fried}}{{Fried}}{1978}]{Fried1978}
{Fried} D.~L.,  1978, Journal of the Optical Society of America (1917-1983),
  68, 1651

\bibitem[\protect\citeauthoryear{Fruchter \& Hook}{Fruchter \&
  Hook}{2002}]{Fruchter2002}
Fruchter A.~S.,  Hook R.~N.,  2002, PASP, 114, 144

\bibitem[\protect\citeauthoryear{{Gillon}, {Lanotte}, {Barman}, {Miller},
  {Demory}, {Deleuil} et~al.,}{{Gillon} et~al.}{2010}]{Gillon2010}
{Gillon} M.,  {Lanotte} A.~A.,  {Barman} T.,  {Miller} N.,  {Demory} B.-O.,
  {Deleuil} M.,    et~al., 2010, A\&A, 511, A3

\bibitem[\protect\citeauthoryear{{Guillot}}{{Guillot}}{2005}]{Guillot2005}
{Guillot} T.,  2005, Annual Review of Earth and Planetary Sciences, 33, 493

\bibitem[\protect\citeauthoryear{{Hardy}}{{Hardy}}{1998}]{Hardy1998}
{Hardy} J.~W.,  1998, {Adaptive Optics for Astronomical Telescopes}

\bibitem[\protect\citeauthoryear{{H{\'e}brard}, {D{\'e}sert}, {D{\'{\i}}az}
  et~al.,}{{H{\'e}brard} et~al.}{2010}]{Hebrard2010}
{H{\'e}brard} G.,  {D{\'e}sert} J.-M.,  {D{\'{\i}}az} R.~F.,    et~al., 2010,
  A\&A, 516, A95+

\bibitem[\protect\citeauthoryear{{Hinkley}, {Oppenheimer}, {Soummer},
  {Sivaramakrishnan}, {Roberts} Jr., {Kuhn}, {Makidon}, {Perrin}, {Lloyd},
  {Kratter} \& {Brenner}}{{Hinkley} et~al.}{2007}]{Hinkley2007}
{Hinkley} S.,  {Oppenheimer} B.~R.,  {Soummer} R.,  {Sivaramakrishnan} A.,
  {Roberts} Jr. L.~C.,  {Kuhn} J.,  {Makidon} R.~B.,  {Perrin} M.~D.,  {Lloyd}
  J.~P.,  {Kratter} K.,    {Brenner} D.,  2007, ApJ, 654, 633

\bibitem[\protect\citeauthoryear{{H{\o}g}, {Fabricius}, {Makarov}, {Urban},
  {Corbin}, {Wycoff}, {Bastian}, {Schwekendiek} \& {Wicenec}}{{H{\o}g}
  et~al.}{2000}]{Hog2000}
{H{\o}g} E.,  {Fabricius} C.,  {Makarov} V.~V.,  {Urban} S.,  {Corbin} T.,
  {Wycoff} G.,  {Bastian} U.,  {Schwekendiek} P.,    {Wicenec} A.,  2000, A\&A,
  355, L27

\bibitem[\protect\citeauthoryear{{Ida} \& {Lin}}{{Ida} \&
  {Lin}}{2004}]{Ida2004}
{Ida} S.,  {Lin} D.~N.~C.,  2004, ApJ, 604, 388

\bibitem[\protect\citeauthoryear{{Juri{\'c}}, {Ivezi{\'c}}, {Brooks}
  et~al.,}{{Juri{\'c}} et~al.}{2008}]{Juric2008}
{Juri{\'c}} M.,  {Ivezi{\'c}} {\v Z}.,  {Brooks} A.,    et~al., 2008, ApJ, 673,
  864

\bibitem[\protect\citeauthoryear{{Kraus} \& {Hillenbrand}}{{Kraus} \&
  {Hillenbrand}}{2007}]{Kraus2007}
{Kraus} A.~L.,  {Hillenbrand} L.~A.,  2007, AJ, 134, 2340

\bibitem[\protect\citeauthoryear{{Law}, {Mackay} \& {Baldwin}}{{Law}
  et~al.}{2006}]{Law2006}
{Law} N.~M.,  {Mackay} C.~D.,    {Baldwin} J.~E.,  2006, A\&A, 446, 739

\bibitem[\protect\citeauthoryear{{Lee} \& {Peale}}{{Lee} \&
  {Peale}}{2002}]{Lee2002}
{Lee} M.~H.,  {Peale} S.~J.,  2002, ApJ, 567, 596

\bibitem[\protect\citeauthoryear{{Lin}, {Bodenheimer} \& {Richardson}}{{Lin}
  et~al.}{1996}]{Lin1996}
{Lin} D.~N.~C.,  {Bodenheimer} P.,    {Richardson} D.~C.,  1996, Nature, 380,
  606

\bibitem[\protect\citeauthoryear{{Liu}, {Burrows} \& {Ibgui}}{{Liu}
  et~al.}{2008}]{Liu2008}
{Liu} X.,  {Burrows} A.,    {Ibgui} L.,  2008, ApJ, 687, 1191

\bibitem[\protect\citeauthoryear{{Mackay}, {Baldwin}, {Law} \&
  {Warner}}{{Mackay} et~al.}{2004}]{Mackay2004}
{Mackay} C.~D.,  {Baldwin} J.,  {Law} N.,    {Warner} P.,  2004, in
  {A.~F.~M.~Moorwood \& M.~Iye} ed., Society of Photo-Optical Instrumentation
  Engineers (SPIE) Conference Series Vol.~5492 of Society of Photo-Optical
  Instrumentation Engineers (SPIE) Conference Series, {High-resolution imaging
  in the visible from the ground without adaptive optics: new techniques and
  results}.
pp 128--135

\bibitem[\protect\citeauthoryear{{Madhusudhan}, {Harrington}, {Stevenson},
  {Nymeyer}, {Campo}, {Wheatley} et~al.,}{{Madhusudhan}
  et~al.}{2011}]{Madhusudhan2011}
{Madhusudhan} N.,  {Harrington} J.,  {Stevenson} K.~B.,  {Nymeyer} S.,  {Campo}
  C.~J.,  {Wheatley} P.~J.,    et~al., 2011, Nature, 469, 64

\bibitem[\protect\citeauthoryear{{Majewski}, {Skrutskie}, {Weinberg} \&
  {Ostheimer}}{{Majewski} et~al.}{2003}]{Majewski2003}
{Majewski} S.~R.,  {Skrutskie} M.~F.,  {Weinberg} M.~D.,    {Ostheimer} J.~C.,
  2003, ApJ, 599, 1082

\bibitem[\protect\citeauthoryear{{Marois}, {Nadeau}, {Doyon}, {Racine} \&
  {Walker}}{{Marois} et~al.}{2003}]{Marois2003}
{Marois} C.,  {Nadeau} D.,  {Doyon} R.,  {Racine} R.,    {Walker} G.~A.~H.,
  2003, in {Mart{\'{\i}}n} E.,  ed., Brown Dwarfs Vol.~211 of IAU Symposium,
  {Differential Simultaneous Imaging and Faint Companions: TRIDENT First
  Results from CFHT}.
p.~275

\bibitem[\protect\citeauthoryear{{Marzari} \& {Nelson}}{{Marzari} \&
  {Nelson}}{2009}]{Marzari2009}
{Marzari} F.,  {Nelson} A.~F.,  2009, ApJ, 705, 1575

\bibitem[\protect\citeauthoryear{{McLaughlin}}{{McLaughlin}}{1924}]{McLaughlin%
1924}
{McLaughlin} D.~B.,  1924, ApJ, 60, 22

\bibitem[\protect\citeauthoryear{{Morton} \& {Johnson}}{{Morton} \&
  {Johnson}}{2011}]{Morton2011}
{Morton} T.~D.,  {Johnson} J.~A.,  2011, ApJ, 729, 138

\bibitem[\protect\citeauthoryear{{Nagasawa}, {Ida} \& {Bessho}}{{Nagasawa}
  et~al.}{2008}]{Nagasawa2008}
{Nagasawa} M.,  {Ida} S.,    {Bessho} T.,  2008, ApJ, 678, 498

\bibitem[\protect\citeauthoryear{{Narita}, {Kudo}, {Bergfors}, {Nagasawa}
  et~al.,}{{Narita} et~al.}{2010}]{Narita2010}
{Narita} N.,  {Kudo} T.,  {Bergfors} C.,  {Nagasawa} M.,    et~al., 2010, PASJ,
  62, 779

\bibitem[\protect\citeauthoryear{{Narita}, {Takahashi}, {Kuzuhara}, {Hirano},
  {Suenaga}, {Kandori}, {Kudo} et~al.,}{{Narita} et~al.}{2012}]{Narita2012}
{Narita} N.,  {Takahashi} Y.~H.,  {Kuzuhara} M.,  {Hirano} T.,  {Suenaga} T.,
  {Kandori} R.,  {Kudo} T.,    et~al., 2012, PASJ, 64, L7

\bibitem[\protect\citeauthoryear{{Pont}, {H{\'e}brard} \& {Irwin}}{{Pont}
  et~al.}{2009}]{Pont2009}
{Pont} F.,  {H{\'e}brard} G.,    {Irwin} J.~M.~o.,  2009, A\&A, 502, 695

\bibitem[\protect\citeauthoryear{{Rasio} \& {Ford}}{{Rasio} \&
  {Ford}}{1996}]{Rasio1996}
{Rasio} F.~A.,  {Ford} E.~B.,  1996, Science, 274, 954

\bibitem[\protect\citeauthoryear{{Rossiter}}{{Rossiter}}{1924}]{Rossiter1924}
{Rossiter} R.~A.,  1924, ApJ, 60, 15

\bibitem[\protect\citeauthoryear{{Schlaufman}}{{Schlaufman}}{2010}]{Schlaufman%
2010}
{Schlaufman} K.~C.,  2010, ApJ, 719, 602

\bibitem[\protect\citeauthoryear{{Schr{\"o}ter}, {Czesla}, {Wolter},
  {M{\"u}ller}, {Huber} \& {Schmitt}}{{Schr{\"o}ter}
  et~al.}{2011}]{Schroter2011}
{Schr{\"o}ter} S.,  {Czesla} S.,  {Wolter} U.,  {M{\"u}ller} H.~M.,  {Huber}
  K.~F.,    {Schmitt} J.~H.~M.~M.,  2011, A\&A, 532, A3

\bibitem[\protect\citeauthoryear{Staley \& Mackay}{Staley \&
  Mackay}{2010}]{Staley2010a}
Staley T.~D.,  Mackay C.~D.,  2010, in Ground-based and Airborne
  Instrumentation for Astronomy III No.~7735 in Proc. SPIE, Data reduction
  strategies for lucky imaging

\bibitem[\protect\citeauthoryear{Stetson}{Stetson}{1987}]{Stetson1987}
Stetson P.~B.,  1987, PASP, 99, 191

\bibitem[\protect\citeauthoryear{{Takeda}, {Kita} \& {Rasio}}{{Takeda}
  et~al.}{2008}]{Takeda2008}
{Takeda} G.,  {Kita} R.,    {Rasio} F.~A.,  2008, ApJ, 683, 1063

\bibitem[\protect\citeauthoryear{{Tokovinin}}{{Tokovinin}}{2002}]{Tokovinin200%
2}
{Tokovinin} A.,  2002, PASP, 114, 1156

\bibitem[\protect\citeauthoryear{{Triaud}, {Collier Cameron}, {Queloz},
  {Anderson}, {Gillon}, {Hebb}, {Hellier}, {Loeillet}, {Maxted}, {Mayor},
  {Pepe}, {Pollacco}, {S{\'e}gransan}, {Smalley}, {Udry}, {West} \&
  {Wheatley}}{{Triaud} et~al.}{2010}]{Triaud2010}
{Triaud} A.~H.~M.~J.,  {Collier Cameron} A.,  {Queloz} D.,  {Anderson} D.~R.,
  {Gillon} M.,  {Hebb} L.,  {Hellier} C.,  {Loeillet} B.,  {Maxted} P.~F.~L.,
  {Mayor} M.,  {Pepe} F.,  {Pollacco} D.,  {S{\'e}gransan} D.,  {Smalley} B.,
  {Udry} S.,  {West} R.~G.,    {Wheatley} P.~J.,  2010, A\&A, 524, A25+

\bibitem[\protect\citeauthoryear{{Tubbs}, {Baldwin}, {Mackay} \& {Cox}}{{Tubbs}
  et~al.}{2002}]{Tubbs2002}
{Tubbs} R.~N.,  {Baldwin} J.~E.,  {Mackay} C.~D.,    {Cox} G.~C.,  2002, A\&A,
  387, L21

\bibitem[\protect\citeauthoryear{{Watson}, {Littlefair}, {Diamond}, {Collier
  Cameron}, {Fitzsimmons}, {Simpson}, {Moulds} \& {Pollacco}}{{Watson}
  et~al.}{2011}]{Watson2011}
{Watson} C.~A.,  {Littlefair} S.~P.,  {Diamond} C.,  {Collier Cameron} A.,
  {Fitzsimmons} A.,  {Simpson} E.,  {Moulds} V.,    {Pollacco} D.,  2011,
  MNRAS, 413, L71

\bibitem[\protect\citeauthoryear{{Weidenschilling} \&
  {Marzari}}{{Weidenschilling} \& {Marzari}}{1996}]{Weidenschilling1996}
{Weidenschilling} S.~J.,  {Marzari} F.,  1996, Nature, 384, 619

\bibitem[\protect\citeauthoryear{{Winn}}{{Winn}}{2010}]{Winn2010a}
{Winn} J.~N.,  2010, {Exoplanet Transits and Occultations}.
pp 55--77

\bibitem[\protect\citeauthoryear{{Winn}, {Fabrycky}, {Albrecht} \&
  {Johnson}}{{Winn} et~al.}{2010}]{Winn2010b}
{Winn} J.~N.,  {Fabrycky} D.,  {Albrecht} S.,    {Johnson} J.~A.,  2010, ApJ,
  718, L145

\bibitem[\protect\citeauthoryear{{Winn}, {Howard}, {Johnson} et~al.,}{{Winn}
  et~al.}{2009}]{Winn2009b}
{Winn} J.~N.,  {Howard} A.~W.,  {Johnson} J.~A.,    et~al., 2009, ApJ, 703,
  2091

\bibitem[\protect\citeauthoryear{{Winn}, {Howard}, {Johnson}, {Marcy},
  {Isaacson}, {Shporer}, {Bakos}, {Hartman}, {Holman}, {Albrecht}, {Crepp} \&
  {Morton}}{{Winn} et~al.}{2011}]{Winn2011}
{Winn} J.~N.,  {Howard} A.~W.,  {Johnson} J.~A.,  {Marcy} G.~W.,  {Isaacson}
  H.,  {Shporer} A.,  {Bakos} G.~{\'A}.,  {Hartman} J.~D.,  {Holman} M.~J.,
  {Albrecht} S.,  {Crepp} J.~R.,    {Morton} T.~D.,  2011, AJ, 141, 63

\bibitem[\protect\citeauthoryear{{Winn}, {Johnson}, {Albrecht}, {Howard},
  {Marcy}, {Crossfield} \& {Holman}}{{Winn} et~al.}{2009}]{Winn2009a}
{Winn} J.~N.,  {Johnson} J.~A.,  {Albrecht} S.,  {Howard} A.~W.,  {Marcy}
  G.~W.,  {Crossfield} I.~J.,    {Holman} M.~J.,  2009, ApJ, 703, L99

\bibitem[\protect\citeauthoryear{{Wu} \& {Murray}}{{Wu} \&
  {Murray}}{2003}]{Wu2003}
{Wu} Y.,  {Murray} N.,  2003, ApJ, 589, 605

\bibitem[\protect\citeauthoryear{{Zacharias}, {Monet}, {Levine}, {Urban},
  {Gaume} \& {Wycoff}}{{Zacharias} et~al.}{2005}]{Zacharias2005}
{Zacharias} N.,  {Monet} D.~G.,  {Levine} S.~E.,  {Urban} S.~E.,  {Gaume} R.,
   {Wycoff} G.~L.,  2005, VizieR Online Data Catalog, 1297, 0

\end{thebibliography}

\end{document}